\documentclass[twocolumn]{aastex61}
\usepackage{CJK}
\usepackage{amsmath}
\usepackage{verbatim}
\definecolor{hlp}{RGB}{50, 80, 255}

\newcommand\aastex{AAS\TeX}

\received{July 9, 2018}
\revised{August 11, 2018}
\accepted{August 24, 2018}

\shorttitle{\aastex\ Infrared Variability of Two Dusty White Dwarfs}
\shortauthors{S. Xu et al.}

\begin{document}
\begin{CJK}{UTF8}{gbsn}
\title{Infrared Variability of Two Dusty White Dwarfs }

\correspondingauthor{Siyi Xu}
\email{sxu@gemini.edu}

\author[0000-0002-8808-4282]{Siyi Xu (许\CJKfamily{bsmi}偲\CJKfamily{gbsn}艺)}
\affil{Gemini Observatory, 670 N. A'ohoku Place, Hilo, HI 96720}

\author{Kate Y. L. Su}
\affiliation{Steward Observatory, University of Arizona, 933 N Cherry Avenue, Tucson, AZ 85721, USA}

\author{Laura Rogers}
\affiliation{Institute of Astronomy, University of Cambridge, Madingley Road, Cambridge CB3 0HA, UK}

\author{Amy Bonsor}
\affiliation{Institute of Astronomy, University of Cambridge, Madingley Road, Cambridge CB3 0HA, UK}

\author{Johan Olofsson}
\affiliation{Instituto de F\'isica y Astronom\'ia, Facultad de Ciencias, Universidad de Valpara\'iso, Av. Gran Breta\~na 1111, Playa Ancha, Valpara\'iso, Chile}
\affiliation{N\'ucleo Milenio Formaci\'on Planetaria - NPF, Universidad de Valpara\'iso, Av. Gran Breta\~na 1111, Valpara\'iso, Chile}

\author{Dimitri Veras}
\affiliation{STFC Ernest Rutherford Fellow}
\affiliation{Centre for Exoplanets and Habitability, University of Warwick, Coventry CV4 7AL, UK}
\affiliation{Department of Physics, University of Warwick, Coventry CV4 7AL, UK}

\author{Rik van Lieshout}
\affiliation{Institute of Astronomy, University of Cambridge, Madingley Road, Cambridge CB3 0HA, UK}

\author{Patrick Dufour}
\affiliation{Institut de Recherche sur les Exoplan$\grave{e}$tes (iREx), Universit$\acute{e}$ de Montr$\acute{e}$al, Montr$\acute{e}$al, QC H3C 3J7, Canada}

\author{Elizabeth M. Green}
\affiliation{Steward Observatory, University of Arizona, 933 N Cherry Avenue, Tucson, AZ 85721, USA}

\author{Everett Schlawin}
\affiliation{Steward Observatory, University of Arizona, 933 N Cherry Avenue, Tucson, AZ 85721, USA}

\author{Jay Farihi}
\affiliation{Department of Physics and Astronomy, University College London, London WC1E 6BT, UK}

\author{Thomas G. Wilson}
\affiliation{Department of Physics and Astronomy, University College London, London WC1E 6BT, UK}
\affiliation{Isaac Newton Group of Telescopes, E-38700 Santa Cruz de La Palma, Spain}

\author{David J. Wilson}
\affiliation{Department of Physics, University of Warwick, Coventry CV4 7AL, UK}
\affiliation{McDonald Observatory, C1400, University of Texas at Austin, Austin, TX 78712, USA}

\author{Boris T. G{\"a}nsicke}
\affiliation{Department of Physics, University of Warwick, Coventry CV4 7AL, UK}

\begin{abstract}

The most heavily polluted white dwarfs often show excess infrared radiation from circumstellar dust disks, which are modeled as a result of tidal disruption of extrasolar minor planets. Interaction of dust, gas, and disintegrating objects can all contribute to the dynamical evolution of these dust disks. Here, we report on two infrared variable dusty white dwarfs, SDSS~J1228+1040 and G29-38. For SDSS~J1228+1040, compared to the first measurements in 2007, the IRAC [3.6] and [4.5] fluxes decreased by 20\% by 2014 to a level also seen in the recent 2018 observations. For G29-38, the infrared flux of the 10~$\mu$m silicate emission feature became 10\% stronger between 2004 and 2007,  We explore several scenarios that could account for these changes, including tidal disruption events, perturbation from a companion, and runaway accretion. No satisfactory causes are found for the flux drop in SDSS~J1228+1040 due to the limited time coverage. Continuous tidal disruption of small planetesimals could increase the mass of small grains and concurrently change the strength of the 10~$\mu$m feature of G29-38. Dust disks around white dwarfs are actively evolving and we speculate that there could be different mechanisms responsible for the temporal changes of these disks.
\end{abstract}

\keywords{circumstellar matter -- minor planets, asteroids -- stars: individual: G29-38, SDSS J122859.93+104032.9 -- white dwarfs.}

\section{Introduction} \label{sec:intro}

G29-38 was the first single white dwarf discovered to display excess infrared emission \citep{ZuckermanBecklin1987}, and follow-up studies have shown that the excess flux arises from a close-in hot dust disk \citep{Graham1990b}. The origin of such a dust disk remained as a mystery until the asteroid tidal disruption model was proposed \citep{DebesSigurdsson2002, Jura2003}. According to this model, the disks are remnants of minor planets that were perturbed into the tidal radius of the white dwarf and eventually became totally disrupted. The infrared excess is often modeled as a geometrically thin and optically thick disk within the tidal radius of the white dwarf \citep{Jura2003}. These compact hot dust disks (temperature $\sim$ 1000~K, size 0.01~au) around white dwarfs are morphologically different from debris disks around main sequence stars (temperature $\sim$ 100~K, size a few tens to hundreds au). There are more than 40 white dwarfs that show infrared excess emission consistent with the presence of dusty disks \citep{Farihi2016}. Some of the dusty white dwarfs also display calcium triplet emission from circumstellar gas that spatially coincide with the dust disk \citep[e.g.][]{Gaensicke2006, Gaensicke2008}. 

About 25--50\% white dwarfs are polluted -- they display elements heavier then helium in their spectra \citep{Zuckerman2003,Zuckerman2010,Koester2014a}. In many cases, continuous accretion onto the white dwarf from circumstellar material is needed due to the short settling times of heavy elements. The connection between atmospheric pollution and dust disks was first explored in \citet{vonHippel2007}. The most heavily polluted white dwarfs are accompanied by an infrared excess from a dust disk. Spectroscopic observations of these polluted atmospheres have opened up a new field of measuring chemical compositions of extrasolar planetary material \citep{JuraYoung2014, Harrison2018, Hollands2018}.

Some polluted white dwarfs are dynamically active and they vary on short timescales. For example, the infrared flux of SDSS~J0959$-$0200 dropped by 35\% between two observations in 2010, and remained at the same level afterwards until at least 2014 \citep{XuJura2014}. The gas emission lines around WD~J1617+1620 disappeared within a few years \citep{Wilson2014}. Most gas disks show gradual variations over a few years \citep{Wilson2015,Manser2016a, Manser2016b, Redfield2017, Dennihy2018}. Recently, transits from an actively disintegrating asteroid were detected around WD~1145+017 \citep{Vanderburg2015} -- a white dwarf that is also heavily polluted, has an infrared excess from a dust disk, and displays absorption lines from circumstellar gas \citep{Xu2016}. The optical light curve of WD~1145+017 is changing on a daily basis, likely due to the vigorous nature of tidal disruption \citep[e.g.][]{Gaensicke2016,Gary2017}.

The dynamical mechanism responsible for white dwarf pollution and tidal disruption is an area of active research \citep[e.g.][]{Veras2016}. The general consensus is that minor planets (i.e. asteroids, comets) and giant planets beyond a few au can survive the post main sequence evolution and orbit around white dwarfs \citep{NordhausSpiegel2013, Mustill2014}. Through different dynamical interactions, e.g. mean motion resonance, planet-planet scattering, secular resonance sweeping, and the Kozai-Lidov effect, the orbits of these minor planets are perturbed -- some are ejected from the system while others can enter into the white dwarf's tidal radius ($\sim$100R$_\mathrm{wd}$, \citealt{Debes2012a, Stephan2017, Mustill2018, Smallwood2018}). In addition, there is evidence for continuous accretion of small planetesimals \citep{Wyatt2014}.

In this paper, we focus on two systems, SDSS~J1228+1040 and G29-38. Their basic parameters are listed in Table~\ref{tab:WDPar}. SDSS~J1228+1040 is the prototype of white dwarfs with circumstellar gas debris \citep{Gaensicke2006}. Its infrared excess was reported in \citet{Brinkworth2009}. Through 12-yr optical spectroscopic monitoring, \citet{Manser2016a} found a gradual variation of the gas emission lines and they proposed it as a result of precession of an asymmetric pattern under general relativity. SDSS~J1228+1040 is also heavily polluted and the composition of the accreting material resembles that of bulk Earth \citep{Gaensicke2012}.

\begin{deluxetable}{lccc}
\tablecaption{White Dwarf Parameters \label{tab:WDPar}}
\tablewidth{0pt}
\tablehead{
\colhead{} & \colhead{SDSS~J1228+1040}  & \colhead{G29-38}
}
\startdata
$T_\mathrm{eff}$ (K)  & 23510 & 11240 \\
$log$ $g$ (cgs)  & 8.16 & 8.00\\
Dom.\tablenotemark{a} & H & H \\
$M_\mathrm{wd}$ ($M_\odot$)  & 0.70 & 0.59\\
$R_\mathrm{wd}$ ($R_\mathrm{\odot}$)   & 0.012 & 0.013\\
$d$ (pc)\tablenotemark{b} & 127 &17.5\\
$V$ (mag)  & 16.2 & 13.0\\
Ref	& \citet{Tremblay2011}& \citet{Subasavage2017}\\
\enddata
\tablenotetext{a}{Dominant element in the white dwarf's atmosphere.}
\tablenotetext{b}{Distance is taken from {\it Gaia DR2} \citep{GaiaDr2}. 
}
\end{deluxetable}

G29-38 was the first white dwarf discovered to have an infrared excess and also the first discovered to display a 10~$\mu$m silicate emission feature \citep{Reach2005a, Reach2009}. The star also has a polluted atmosphere \citep{Koester1997} and recent HST/COS observations show that it is accreting from volatile-poor material that is similar in composition to the bulk Earth \citep{Xu2014}. G29-38 is among the first variable white dwarfs discovered \citep{ShulovKopatskaya1974,McGrawRobinson1975}. The newest addition to the wonders of G29-38 comes from the discovery of molecular hydrogen in its atmosphere, which provides an additional constraint of its stellar parameters \citep{Xu2013b}. It is among the hottest stellar environments with a molecular hydrogen detection.

Here, we report infrared observations of the dust disks around SDSS~J1228+1040 and G29-38, demonstrating for the first time that these two disks are intrinsically variable. The rest of the paper is organized as follows. Observation and data reduction are presented in Section~\ref{sec:data}. In Section~\ref{sec:model}, we present some disk models that could explain the temporal variations of the infrared luminosity. Possible scenarios are explored in Section~\ref{sec:interp} and results are summarized in Section~\ref{sec:conclusion}.

\section{Observation And Data Reduction \label{sec:data}}

\subsection{Spitzer/IRAC and MIPS Observations \label{sec:spitzerphot}}

SDSS~J1228+1040 and G29-38 have been observed a few times with Spitzer/IRAC \citep{Fazio2004}, as summarized in Tables~\ref{tab:spitzer2} and \ref{tab:spitzer1}. Both stars are well detected in each frame and separated from background stars. We performed aperture photometry on individual artifact-corrected CBCD (Corrected Basic Calibrated Data) frames with the IDL programs BOX$\_$CENTROIDER.PRO and APER.PRO. An aperture radius of 3 native pixels (3{\farcs}6) and a sky annulus of 12-20 pixels (14{\farcs}4 - 24{\farcs}0) were used for the analysis. Aperture correction, pixel phase, and array location correction were also performed. For each epoch, we report the average flux weighted by the signal-to-noise ratio of each measurement and take the weighted uncertainty as the final uncertainty. Since we are interested in relative flux difference, our reported numbers only include measurement uncertainty. As an additional check, we also performed aperture photometry on the combined mosaic and obtained similar results.

\begin{deluxetable*}{lllllcccccc}
\tablecaption{Spitzer Fluxes of SDSS~J1228+1040\label{tab:spitzer2}}
\tablewidth{0pt}
\tablehead{
\colhead{UT Date} & \colhead{MJD} & \colhead{PID} & \colhead{AORKEY} & \colhead{Time}  & \colhead{3.6 $\mu$m } & \colhead{4.5 $\mu$m} &  \colhead{5.8 $\mu$m } & \colhead{8.0 $\mu$m } & \colhead{MIPS}\\
 & & & & \colhead{(sec)} & \colhead{($\mu$Jy)}  & \colhead{($\mu$Jy)} & \colhead{($\mu$Jy)} & \colhead{($\mu$Jy)}& \colhead{($\mu$Jy)}
}
\startdata
2007 Jun 30	& 53281.3	& 40048	& 22247936	& 100$\times$10	& 228$\pm$10 & 235$\pm$12 & 225$\pm$12 & 246$\pm$31 & ... 	 \\
2014 Sept 3 	& 56903.4 &10032	& 49253376&	30$\times$30	& 184$\pm$9 	& 188$\pm$8	&...	& ... & ...\\
2018 May 9	& 58247.6	& 13216	& 64912384	& 12$\times$10 & 180$\pm$9 & 195$\pm$10 & ... & ... & ... \\
\\
2008 Jul 25	& 54672.1	&	50118	&25459712, 25459456, 25459200	&	10$\times$60 &...	&	...& ...& ...& 24 $\mu$m: 129$\pm$10\\
\enddata
\end{deluxetable*}

\begin{deluxetable*}{lllllcccccc}
\tablecaption{Spitzer Fluxes of G29-38\label{tab:spitzer1}}
\tablewidth{0pt}
\tablehead{
\colhead{UT Date} & \colhead{MJD} & \colhead{PID} & \colhead{AORKEY} & \colhead{Time\tablenotemark{a}}  & \colhead{3.6 $\mu$m } & \colhead{4.5 $\mu$m} &  \colhead{5.8 $\mu$m } & \colhead{8.0 $\mu$m } & \colhead{MIPS }\\
 & & & & \colhead{(sec)} & \colhead{(mJy)}  & \colhead{(mJy)} & \colhead{(mJy)} & \colhead{(mJy)}& \colhead{(mJy)} 
}
\startdata
2004 Nov 26	&53335.5	& 2313,3548	& 10119424, 11124224& 30$\times$20/26\tablenotemark{b} &  8.11$\pm$0.12 & 8.85$\pm$0.10 & 8.22$\pm$0.12& 8.20$\pm$0.10 & ...\\
2005 Dec 22	&53727.0	& 20026 &13835264 &12$\times$10  & 7.91$\pm$0.17 & ... & 8.28$\pm$0.23 & ...& ...\\
2007 Dec 26	&54460.5	& 40369 & 22957312 & 30$\times$5 & 8.01$\pm$0.22 & 8.89$\pm$0.22 & 8.13$\pm$0.22 & 8.38$\pm$0.20& ...\\
2007 Dec 28	& 54462.2	& 40369 & 22960896 & 30$\times$5 & 8.17$\pm$0.23 & 8.99$\pm$0.22	& 8.25$\pm$0.23 & 8.55$\pm$0.20& ...\\
2007 Dec 31	& 54465.5 & 40369 & 22961152 & 30$\times$5 & 8.35$\pm$0.23 & 8.98$\pm$0.22 & 8.44$\pm$0.23 & 8.48$\pm$0.20 & ...\\
2008 Jul 18	&54665.7 & 40369 & 22961408& 30$\times$6   & 8.04$\pm$0.22 & 9.38$\pm$0.22 & 8.38$\pm$0.23 & 8.84$\pm$0.21& ...\\
2009 Aug 19	& 55062.6 &60161 & 35008512 & 30$\times$21 & ... & 8.99$\pm$0.12 & ... & ...& ... \\
\\
2004 Dec 02 & 53341.2 & 2313 & 10149376 & 10$\times$3 &... &... & ...&... & 24 $\mu$m: 2.75$\pm$0.05  \\
2008 Jul 28 & 54675.3 & 50401& 26134016& 10$\times$20& ...& ...&... &... & 70 $\mu$m: $<$ 5.1 (3$\sigma$)\\
\enddata
\tablenotetext{a}{The first value is frame time in seconds and second value is the number of frames.}
\tablenotetext{b}{20 frames for 3.6~$\mu$m and 5.8~$\mu$m and 26 frames for 4.5~$\mu$m and 8.0~$\mu$m.}
\end{deluxetable*}

\begin{figure*}
\gridline{\fig{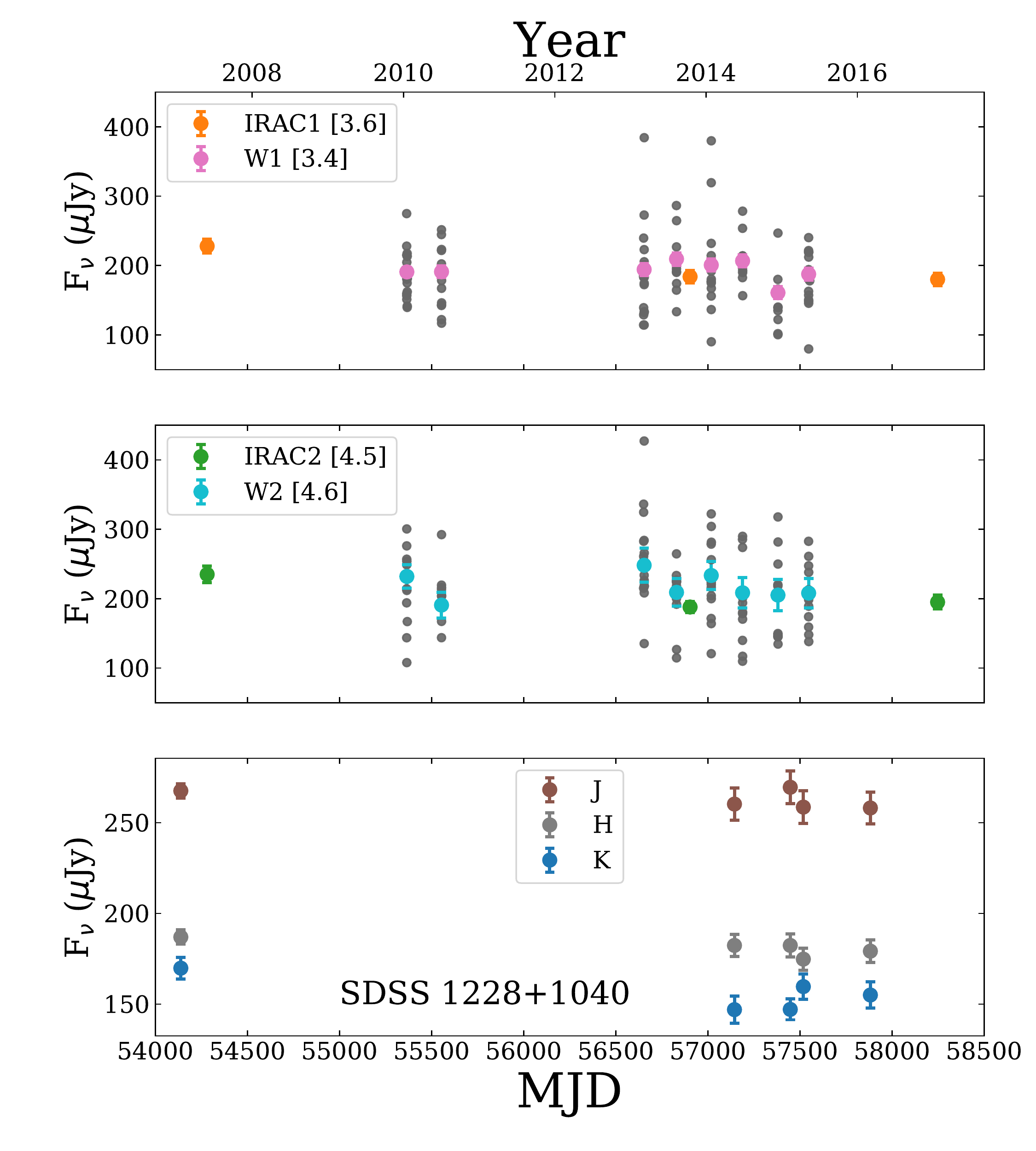}{0.5\textwidth}{}
          \fig{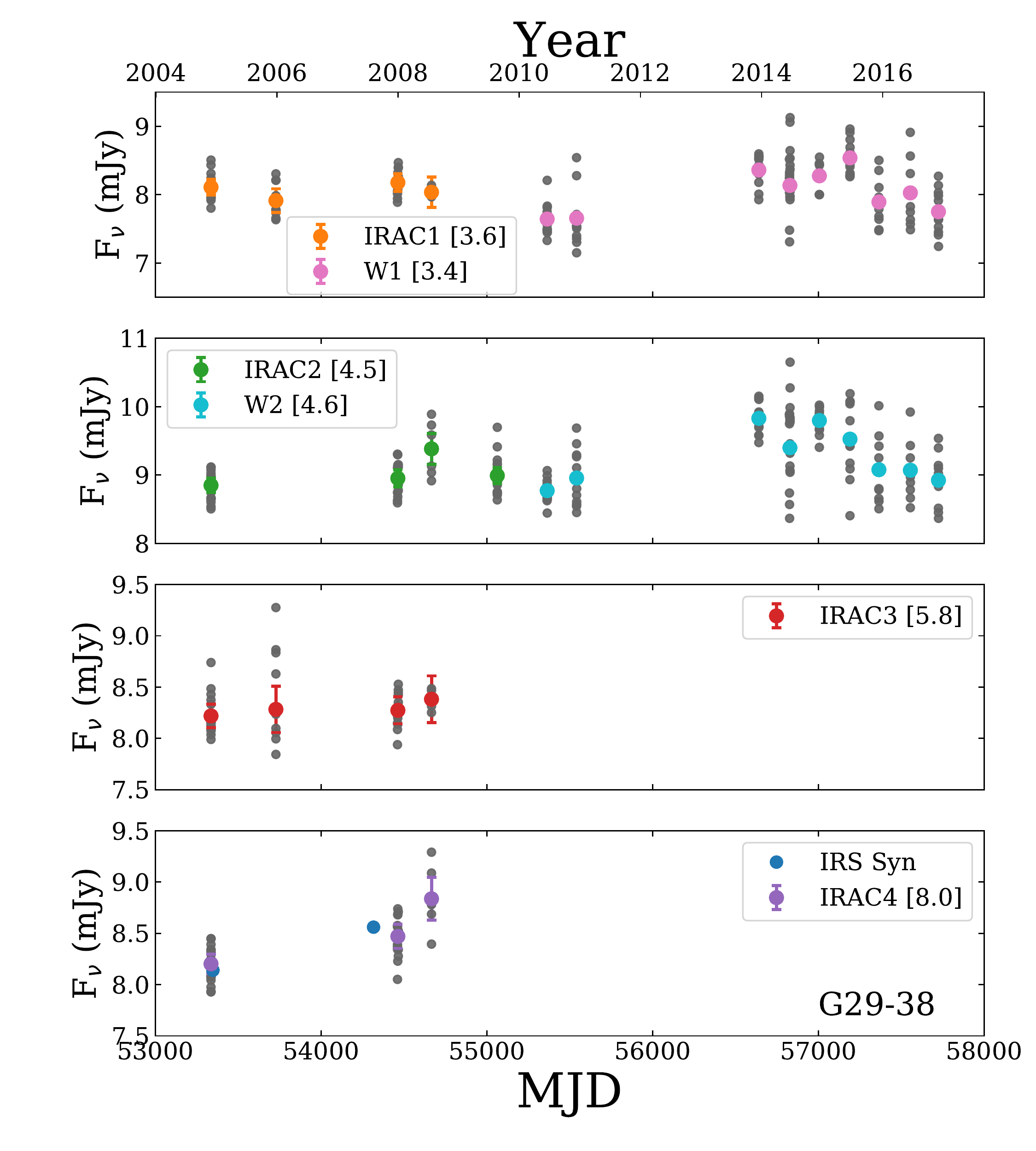}{0.5\textwidth}{}
}
\caption{Infrared fluxes of SDSS~1228+1040 and G29-38. The grey dots represent individual measurements while the colored dots with error bars represent the average flux. For SDSS~1228+1040, a 20\% (3$\sigma$) flux drop is detected between the first two IRAC epochs and a 13\% (2.7$\sigma$) drop is detected at K band between the first two UKIRT observations. For G29-38, there are some low level variabilities. The synthetic photometry from IRS spectra are also presented at the [8.0] panel.
\label{fig:spitzer}}
\end{figure*}

For SDSS~J1228+1040, we found significant variability in the first two {\it Spitzer} observations and the flux levels have remained the same in the third epoch, as shown in Fig.~\ref{fig:spitzer}. The infrared fluxes have dropped by 20\% at 3$\sigma$ significance at both [3.6] and [4.5]. For the background stars of similar brightness, we found their fluxes agree to within 1\% for both [3.6] and [4.5] at all three epochs. The relative flux drop of SDSS~J1228+1040 between 2007 and 2014 is detected at high significance.

For G29-38, there is some dispersion in the flux level in the individual CBCD frames, as indicated by the spread of the grey dots in Fig.~\ref{fig:spitzer}. The maximum flux difference between different epochs/AORs is 3.3\% (1.2$\sigma$) at [3.6], 6.0\% (2.1$\sigma$) at [4.5], 2.0\% (0.6$\sigma$) at [5.8], and 7.7\% (2.8$\sigma$) at [8.0]. As a sanity check, we performed aperture photometry on a few background stars and found that their fluxes agree within 2\%, 2\%, 6\%, and 7\% for [3.6], [4.5], [5.8] and [8.0] respectively. G29-38 is the brightest star in the field of view and yet its photometry stability is worse compared to the faint field stars, particularly at [3.6] and [4.5]. This is expected because the exposure time is 30~sec, which is much shorter than the pulsation periods (typically tens of minutes, see \citealt{Kleinman1998}). We defer the discussion of pulsation induced flux variation to Section~\ref{sec:g29-38pul}. 
 
{\it Spitzer}/MIPS observations \citep{Rieke2004} of SDSS~J1228+1040 and G29-38 have been published by using early {\it Spitzer} pipelines and calibrations \citep{Brinkworth2009,Reach2009,Farihi2014}. Here, we reprocessed all the data using the MIPS instrument in-house pipeline with the final calibration established for the mission (described in \citealt{Sierchio2014}). SDSS~J1228+1040 has three deep 24~$\mu$m observations with a total exposure time of 600 s. The source is weakly detected in the individual mosaic. G29-38 is a clean point-like source at 24 $\mu$m but not detected at 70~$\mu$m. We measure the photometry from the mosaics, and then adopt the average and weighted uncertainty as the final flux and the uncertainty, respectively, as reported in Table~\ref{tab:spitzer2}.

\subsection{WISE Observations}

Since the launch of the Wide-field Infrared Survey Explorer (WISE; \citealt{Wright2010}) in 2010, SDSS~J1228+1040 and G29-38 have been observed by WISE and now NEOWISE \citep{Mainzer2011} every half year. Because of their faintness, these stars are only detected in the two shortest bands in WISE, which have similar bandwidths as IRAC [3.6] and [4.5]. By using the NASA/IPAC infrared science archive, we extracted the photometry from the multi-epoch photometry table and the Single Exposure Source Table for WISE and NEOWISE, respectively. We calculated the weighted average flux for all observations taken within ten days, as plotted in Fig.~\ref{fig:spitzer}. For SDSS~J1228+1040, the uncertainties are too large to detect the 20\% flux drop identified in IRAC. For G29-38, similar to the IRAC observations, the WISE data show that G29-38 is not photometrically stable and there are some low level variabilities from pulsation.

\subsection{Spitzer/IRS Spectroscopy}

G29-38 has also been observed with the {\it Spitzer}/IRS instrument \citep{Houck2004} during the cryogenic mission: 2004 December 8 (AORKEY
10184192), 2006 June 30 (AORKEY 13828096) and 2007 August 4 (AORKEY
22957568). The Short-Low (SL, 5.2--14.5~$\mu$m, resolution $\sim$ 100) module was used in
both 2004 and 2007, while the SL2 (5.2--8.7~$\mu$m) and
Long-Low (LL, 19.4--38~$\mu$m, resolution $\sim$ 100) modules were used in 2006. The 2004 and
2006 IRS spectra that were reduced by an early version of the pipeline were
published by \citet{Reach2009}. For this study, we used the spectra from
the CASSIS website that provides uniform, high-quality IRS spectra
optimally extracted for point-like sources \citep{Lebouteiller2011}.

\begin{figure}
\epsscale{1.15}
\plotone{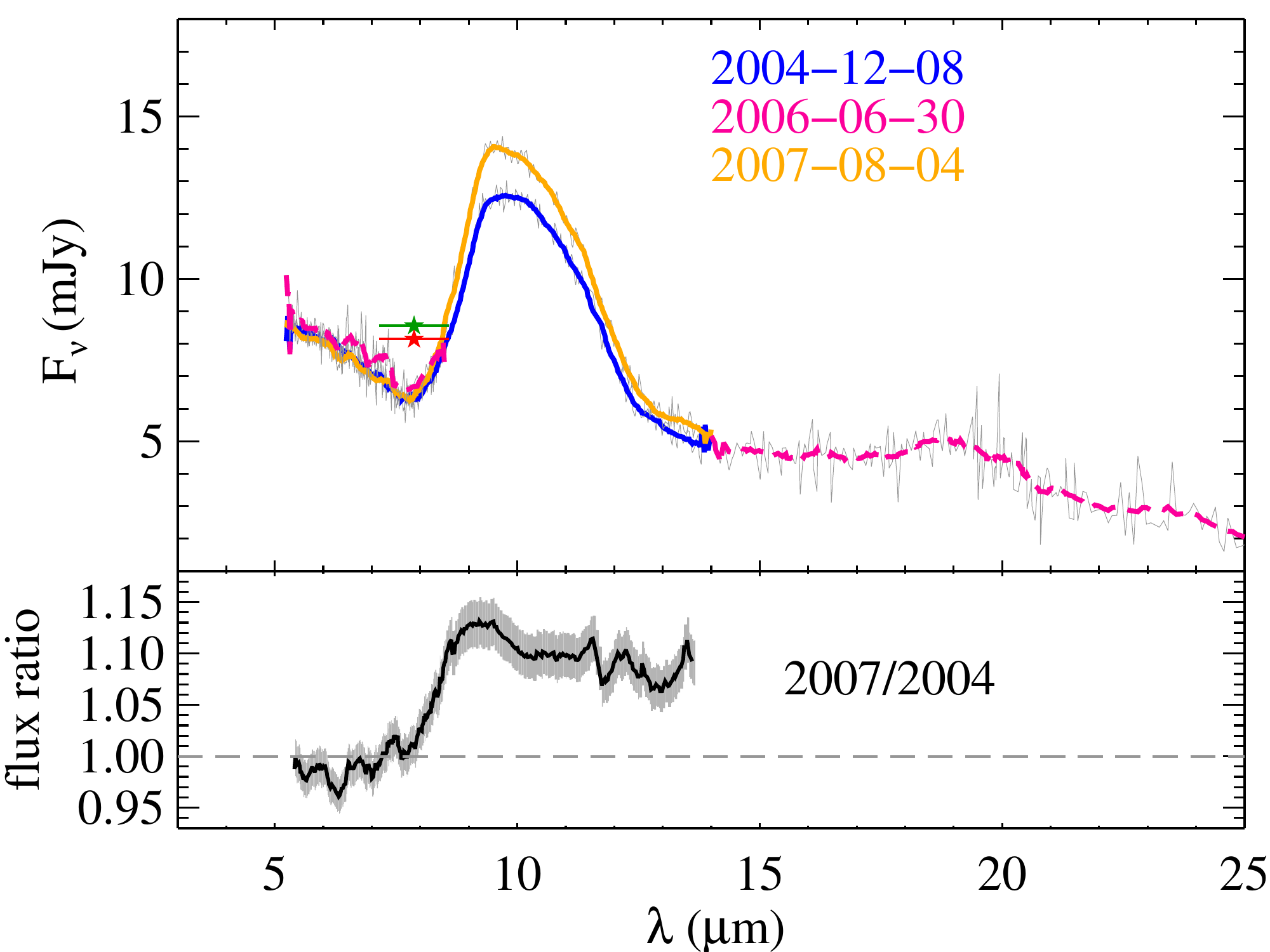}.
\caption{Upper panel shows the smoothed IRS 5--15 $\mu$m spectra for G29-38 taken in 2004 (blue, 5.2--14.5~$\mu$m), 2006 (pink, 5.2--8.7~$\mu$m and 19.4--38~$\mu$m), and 2007 (yellow, 5.2--14.5~$\mu$m). For comparison, the unsmoothed spectra are also shown as thin grey lines. The [8.0] IRS synthesized photometry is shown as the star symbols (red for 2004, and dark green for 2007) with horizontal bars representing the half bandpass. The bottom panel shows the flux ratio between 2004 and 2007.
\label{fig:g29-38irs}}
\end{figure}

As shown in Fig.~\ref{fig:g29-38irs}, the flux of G29-38 in the 5--7~$\mu$m region agrees
within 2\% over the three-year period, but the 2007 flux in the 10~$\mu$m silicate feature region increased by 10\%. The temporal
variability in the IRS spectra is in line with the IRAC photometry
presented in Section \ref{sec:spitzerphot}. To make a direct comparison, we computed
the synthesized [8.0] photometry using the observed IRS
spectra: 8.14 mJy in 2004 and 8.56 mJy in 2007, consistent with the 5\%
increase in the [8.0] photometry from 2004 to 2007 (see Table~\ref{tab:spitzer1}). Given that the 5--7 $\mu$m IRS flux between the 2004 and 2007 epochs agrees within 2\%, we conclude that
the temporal variability between 2004 and 2007 IRS spectra is significant.

\subsection{UKIRT JHK Observations \label{sec:jhkphot}}

Since 2015, we have been monitoring dusty white dwarfs with the UK Infrared Telescope (UKIRT) to understand the origin of their variability. Details of the survey strategy and data reduction will be presented in {\it Rogers et al. (in prep)}. For SDSS~J1228+1040 and G29-38, the raw data were processed with pipelines produced by the Cambridge Astronomical Survey Unit (CASU). We compare the standard deviation of each frame and the uncertainty of the average flux and take the larger one of the two values as the measurement uncertainty, as listed in Table~\ref{tab:ukirt2} and Tables~\ref{tab:ukirt1}. Calibration uncertainties are not considered here. For SDSS~J1228+1040, we detected a 13\% (2.7$\sigma$) drop in the K band flux between 2007 and 2015 and it has remained at the same level since then. As shown in Fig.~\ref{fig:spitzer}, this K band flux drop is likely related to the IRAC flux change. For G29-38, there is a 7\% (3.7$\sigma$) variation in the K band and it is consistent with pulsation induced infrared variation (see Section~\ref{sec:g29-38pul}). 

\begin{deluxetable*}{lcccl}
\tablecaption{Near-infrared Photometry of SDSS~J1228+1040\label{tab:ukirt2}}
\tablewidth{0pt}
\tablehead{
\colhead{UT Date} & \colhead{J (mag)} & \colhead{H (mag)} &  \colhead{K (mag)} & \colhead{Ref}
}
\startdata
2007 Feb 7	& 16.893 $\pm$ 0.016 & 16.841 $\pm$ 0.023  & 16.425 $\pm$ 0.038		& UKIDSS\\
2015 May 2	& 16.923 $\pm$ 0.017 & 16.868 $\pm$ 0.017	& 16.582 $\pm$ 0.045	& UKIRT\\
2016 Feb 29	& 16.885 $\pm$ 0.016	& 16.868 $\pm$ 0.020	& 16.581 $\pm$ 0.027	& {\it this work}\\
2016 May 10	& 16.930 $\pm$ 0.020	& 16.914 $\pm$ 0.019	& 16.492 $\pm$ 0.034	& {\it this work}\\
2017 May 9	& 16.932 $\pm$ 0.018	& 16.887 $\pm$ 0.019	& 16.524 $\pm$ 0.039	& {\it this work}\\ 
\enddata
\end{deluxetable*}

\begin{deluxetable*}{lcccl}
\tablecaption{Near-infrared Photometry of G29-38\label{tab:ukirt1}}
\tablewidth{0pt}
\tablehead{
\colhead{UT Date} & \colhead{J (mag)} & \colhead{H (mag)} &  \colhead{K (mag)} & \colhead{Ref}
}
\startdata
2000 Aug 7 & 13.132 $\pm$ 0.026 & 13.075 $\pm$ 0.022 & 12.689 $\pm$ 0.029 & 2MASS \\
2009	 Jun 21& 13.123 $\pm$ 0.002 &  13.025 $\pm$ 0.002	& 	12.552 $\pm$ 0.003 & UKIDSS\\
2015 Aug 9	&	13.127 $\pm$ 0.010 & 13.038 $\pm$ 0.010	& 12.595 $\pm$ 0.013 & {\it this work}\\
2016 Jul 6	&	13.141 $\pm$ 0.010 & 13.042 $\pm$ 0.010	& 12.623 $\pm$ 0.020 & {\it this work}\\
\enddata
\end{deluxetable*}

\subsection{Optical Photometric Monitoring \label{sec:wd1226opt}}

SDSS~J1228+1040 was observed on 2018 March 21 (UT) using the University of
Arizona's 61 inch (1.55 m) Kuiper telescope on Mt.\ Bigelow, Arizona. The camera was equipped 
with the Mont4k CCD, binned 3$\times$3 to 0\farcs43 per pixel. The Schott-8612 filter (a broad
white-light filter) was used. We adopted an exposure time of 30~sec, resulting in a cadence of 44.8~sec including set-up and readout. 
Conditions during the observation were nearly photometric and moonless. We obtained a continuous 6~hr observations of SDSS~J1228+1040 and reached a typical S/N of 300 per frame.

\begin{figure}[ht!]
\epsscale{1.15}
\plotone{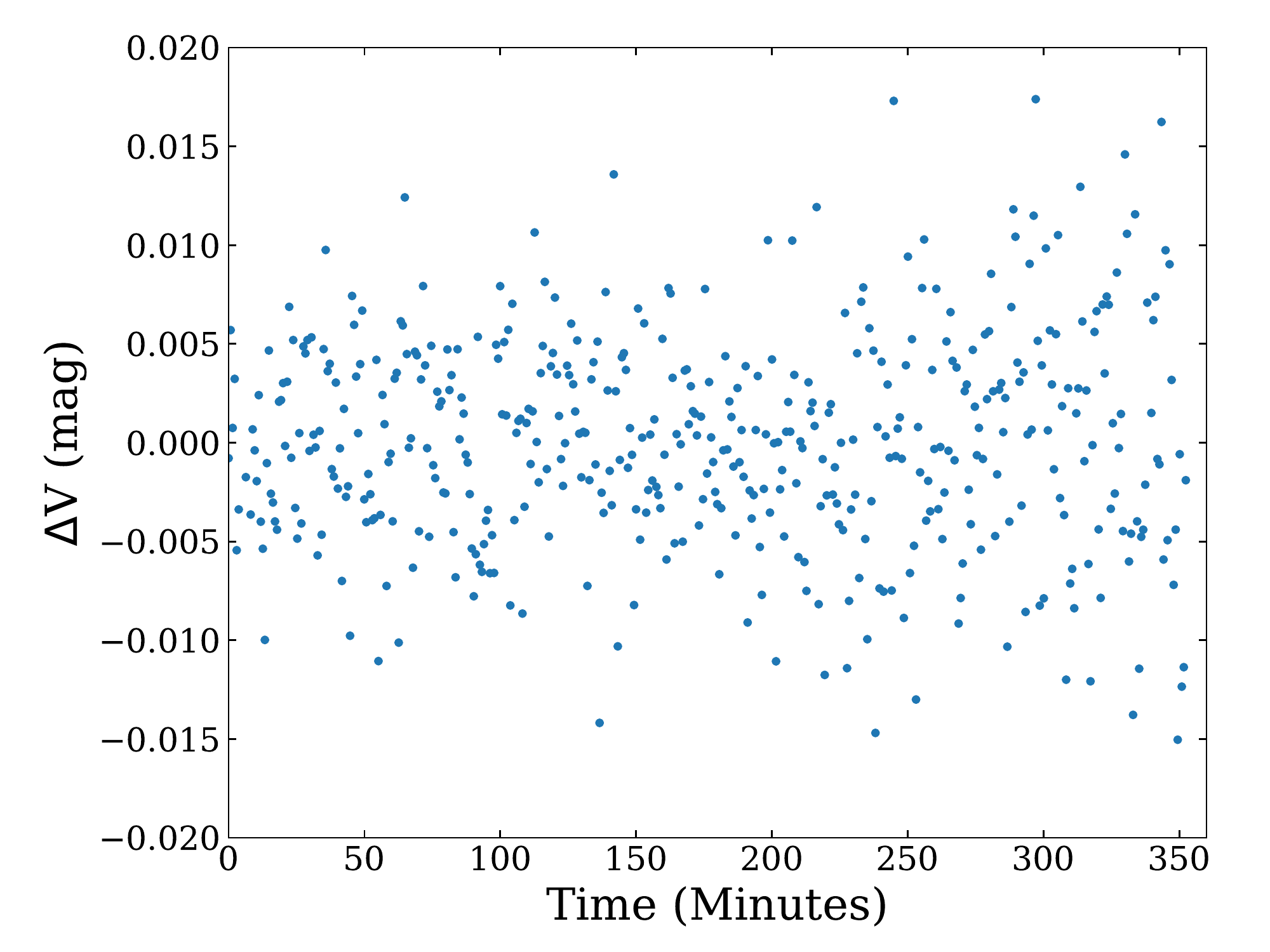}
\caption{Optical monitoring of SDSS~J1228+1040 from the 61 inch telescope. No optical variability is detected and the standard deviation of the light curve is 0.006~mag. The scatter towards the end of the observations is due to the increasing airmass of the target.
\label{fig:wd1226_61}}
\end{figure}

All the images were bias-subtracted, flat-fielded, and bad
pixel-cleaned in the usual manner. Aperture photometry was performed
using the task PHOT in the IRAF DAOPHOT
package. We performed relative photometry by referencing to other 8
nearby stars within the 10\arcmin$\times$10\arcmin\ field of
view. After correcting the color response due to the airmass between
our target (blue star) and field stars (generally red stars), we found no significant
optical variability with a standard deviation of 0.006~mag (see Fig.~\ref{fig:wd1226_61}).

\section{Disk Modeling \label{sec:model}}

\subsection{SDSS~J1228+1040: Opaque Disk \label{sec:wd1226model}}

\begin{figure*}
\centering
\includegraphics[width=0.6\textwidth]{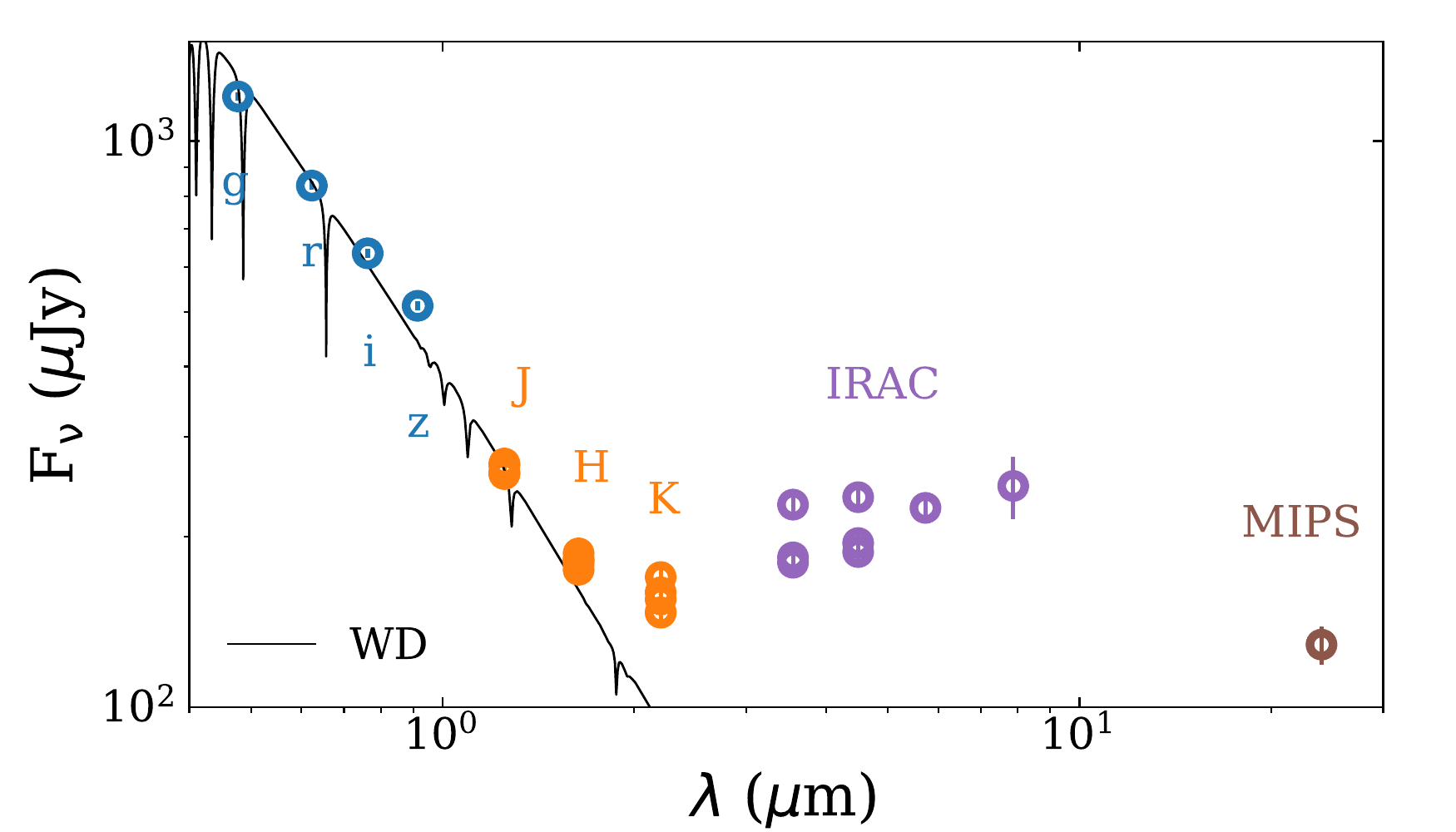}
\includegraphics[width=0.6\textwidth]{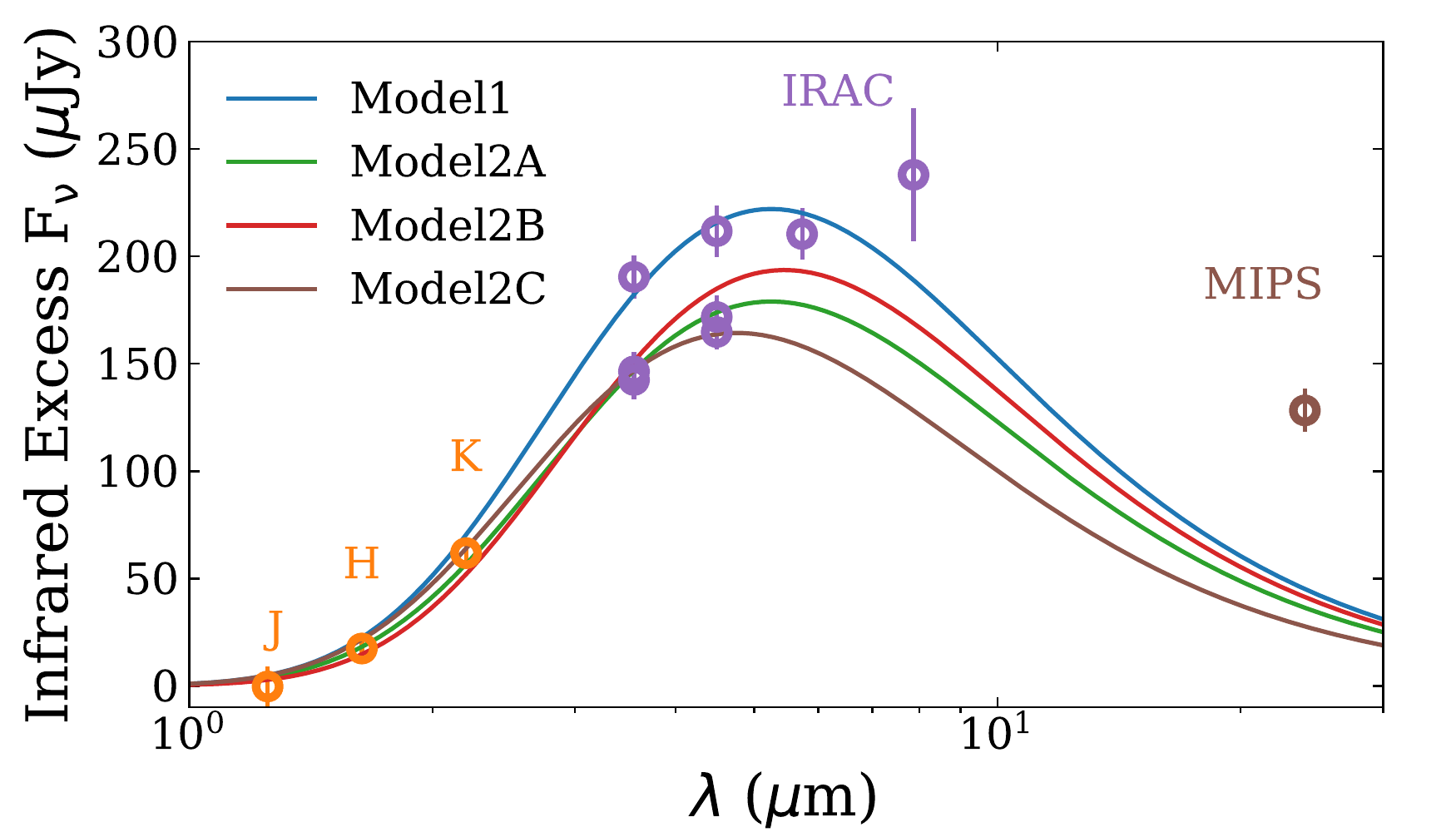}
\caption{SED fits for SDSS~J1228+1040. The photometry points are from the {\it SDSS} (griz), UKIRT (JHK), and Spitzer (IRAC and MIPS). The top panel shows the entire SED while the bottom panel is the excess infrared flux. The parameters for the models are listed in Table~\ref{tab:wd1226fit}. The change of the infrared fluxes can either be modeled by a change of the inclination, inner disk radius, or outer disk radius.
\label{fig:wd1226SED}}
\end{figure*}

The white dwarf SDSS~J1228+1040 is stable to 0.006~mag at optical wavelengths (see Section~\ref{sec:wd1226opt}) and therefore, the infrared variability must come from the circumstellar material. In terms of a simple geometrically thin and optically thick disk model \citep{Jura2003}, we assume that the gas and dust occupy a similar region, and that sublimation and/or collisions of dust particles feed material into the gas disk, which then get accreted by the star. Its circumstellar gaseous disk is inferred to have an inclination of 70$^{\circ}$, a small eccentricity, and a radius of 60--120 R$_\mathrm{wd}$ \citep{Gaensicke2006}. To minimize the number of free parameters, we take the disk eccentricity to be 0. There are three free parameters, inner disk radius R$_\mathrm{in}$, outer disk radius R$_\mathrm{out}$, and line-of-sight disk inclination $i$. The white dwarf flux was calculated as with parameters listed in Table~\ref{tab:WDPar} \citep{Dufour2017}. To match the SDSS $rgiz$ photometry, an additional scaling factor of 0.92 is applied. To calculate the excess infrared flux, we subtract the white dwarf flux from the measured flux, assuming a 2\% uncertainty in the white dwarf flux, as shown in Fig~\ref{fig:wd1226SED}.

For the fit, we consider both the IRAC fluxes and the average value of the JHK photometry, since there are no significant variations. For the 2007 observation, we fix the inclination at 70$^{\circ}$, which is the same as the gas debris. We performed a chi squared minimization and the best fit parameters are listed in Table~\ref{tab:wd1226fit}. With all four IRAC points, the disk parameters are well constrained. However, for the 2014 and 2018 data, there are many parameters that could fit the data given the lack of longer wavelength observations. So we started with the best-fitted parameters for the 2007 data and vary one parameter at a time. The change of the infrared flux can be modeled by a decrease of the total emitting area, either by a change of disk inclination (model 2A) or a change of the disk radius and correspondingly a change in dust temperature (models 2B and 2C). The opaque disk model cannot match the high MIPS flux, either because the disk flux has changed between the IRAC and MIPS observations or the presence of an optically thin layer, similar to the case for G29-38, as discussed in the next section.

\begin{deluxetable}{lccccl}
\tablecaption{ Fitting Parameters for SDSS~J1228+1040 \label{tab:wd1226fit}}
\tablewidth{0pt}
\tablehead{
& \colhead{Model 1}	& \colhead{Model 2A}	& \colhead{Model 2B} & \colhead{Model 2C}
}
\startdata
inner radius R$_\mathrm{in}$ & 27R$_\mathrm{wd}$	& 267$_\mathrm{wd}$	&  30R$_\mathrm{wd}$	& 27R$_\mathrm{wd}$\\
outer radius R$_\mathrm{out}$ & 63R$_\mathrm{wd}$	&63R$_\mathrm{wd}$	& 63R$_\mathrm{wd}$	&49R$_\mathrm{wd}$\\
inclination $i$ & 70$^\circ$ & 74$^\circ$ & 70$^\circ$& 70$^\circ$\\
$\chi^2$ & 7.2	& 2.7& 10.6	& 0.8\\
\enddata
\tablecomments{Model 1 is the best fit parameters for the 2007 IRAC data ([3.6], [4.5], [5.8], and [8.0]), while Model 2A, 2B, and 2C are for the 2014 and 2018 IRAC data ([3.6] and [4.5]).
}
\end{deluxetable}

\citet{Manser2016a} found that the multi-year variation of the emission lines is consistent with precession of an eccentric pattern under general relativity. Note that such precession could not explain the infrared variability assuming the disk is geometrically thin and optically thick. However, if the disk is not geometrically thin, as suggested by recent studies \citep{KenyonBromley2017a}, the infrared flux change might be explained by the obscuring of dust materials from a different part of the disk. This model is beyond the scope of current work and will be explored in a future study.

\subsection{G29-38 \label{sec:g29-38disk}}

\subsubsection{Stellar Pulsation
\label{sec:g29-38pul}}

Through 10-yr optical photometric monitoring of G29-38,
\citet{Kleinman1998} identified 19 pulsation modes with periods
between 100 and 1300 sec, and not all the modes are excited at the same
time. Although the white dwarf pulsation is negligible in the
infrared, the flux of the dust disk would change as it reprocesses the
star light. To study the effect of white dwarf pulsation on the dust
disk, time-series infrared observations of G29-38 have been performed
\citep{Graham1990b, Reach2009}. A 190~sec period was identified with an
amplitude of 2.5\% at K band and 4\% at [3.6]; interestingly, this
period is not detected in the simultaneous optical light curve. Both
\citet{Graham1990b} and \citet{Reach2009} postulate that the 190~sec
period of the dust disk is induced from pulsations with temperature
changes along the latitude (e.g. m=0 modes), which has a net
temperature change on the dust disk; while the other modes are
confined to regions perpendicular to the dust disk (e.g. sectoral, m =
l modes) and they cause no net temperature change on the dust disk.

The largest pulsation amplitude is $\pm$5\% in the optical,
which corresponds to a 2\% white dwarf temperature change
\citep{Kleinman1998}. The average stellar temperature is 11240 K
(see Table~\ref{tab:WDPar}), suggesting that the stellar temperature can be as low as 11015~K in the low state, but as high as 11465~K in the high state. Because
the disk is directly heated by the white dwarf, the temporal change of
the 10~$\mu$m feature might be caused by the different stellar
temperatures due to pulsation. In the following subsection, we will
first introduce our simple two-component model that can fit the disk
SED, and then explore the likely change in the observed SED due to two
different states of stellar heating.

\subsubsection{A Two-Component Disk Model \label{sec:g29-38model}}

\begin{figure*}[ht!]
\gridline{\fig{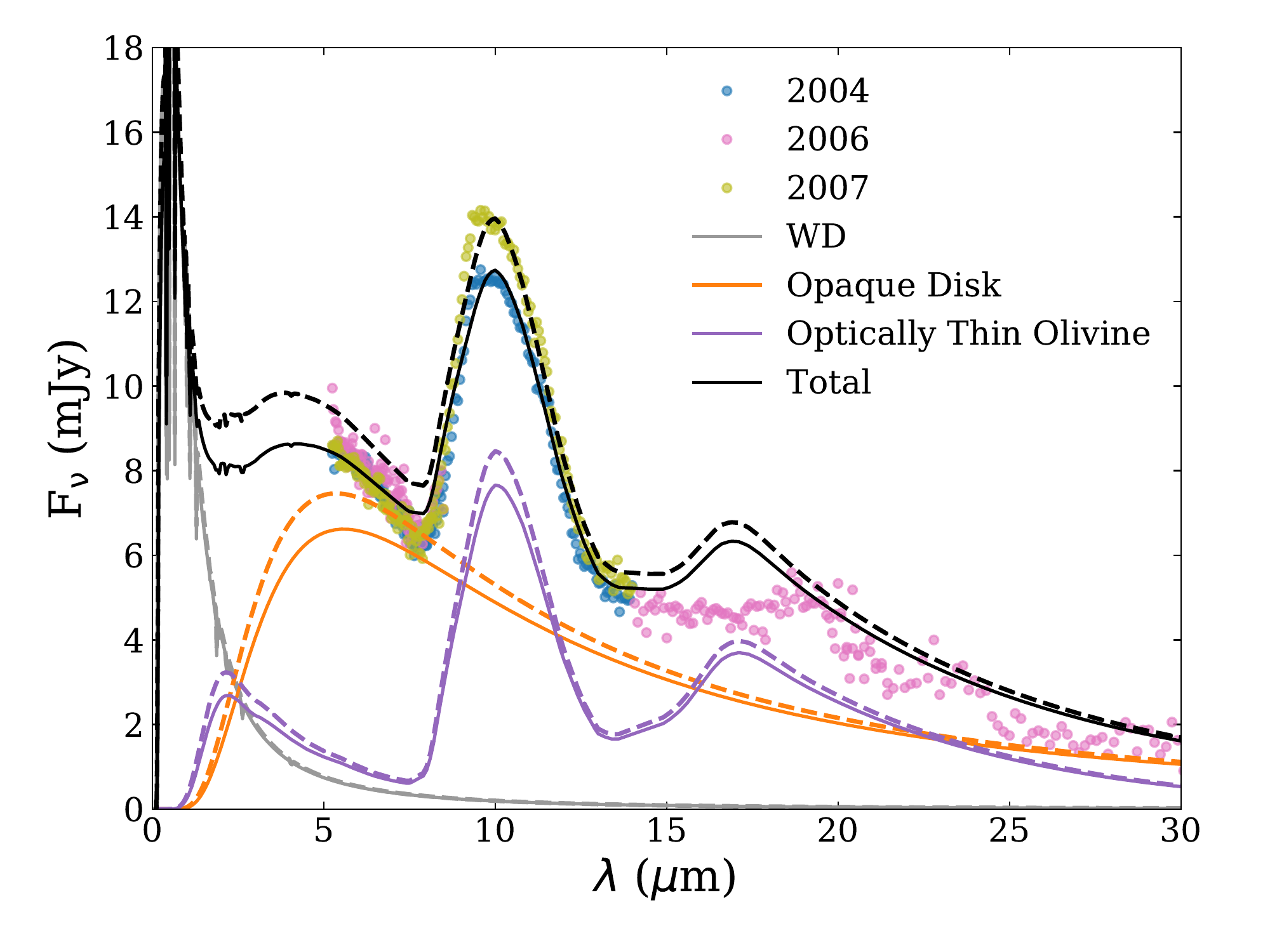}{0.5\textwidth}{}
\fig{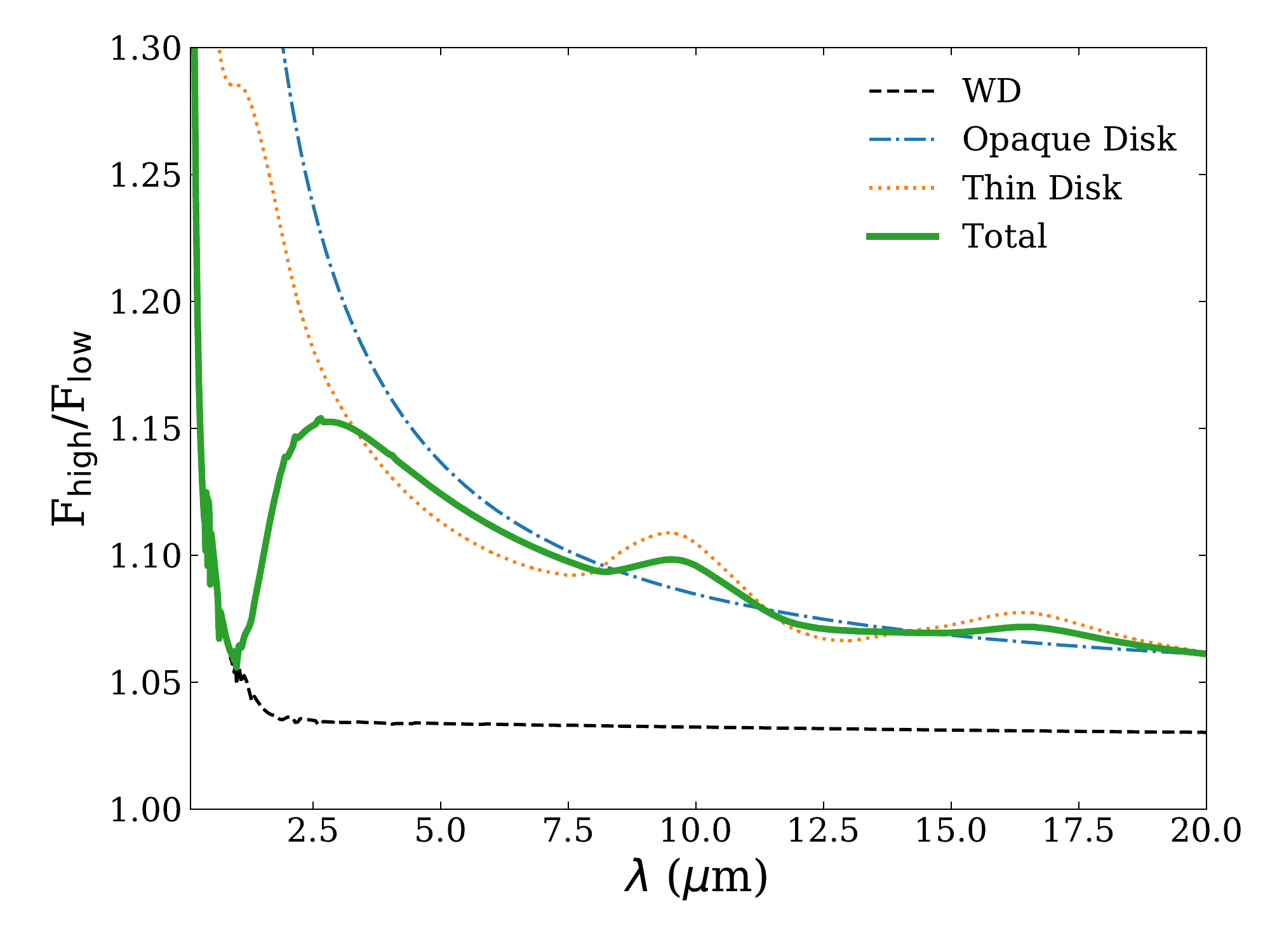}{0.5\textwidth}{} }
\caption{Two-component disk model with T$_\mathrm{wd,low}$=11015 K and T$_\mathrm{wd,high}$=11465 K. {\it Left panel}: The dashed line represents the high
temperature state while the solid line represents the low temperature
state. The stellar temperature is the only difference and the disk parameters are all kept the same. {\it Right panel}: The black, blue, orange and green lines
represents the ratio of the highest and lowest flux for the white
dwarf, optically thick disk, optically thin disk, and the total flux,
respectively. If the 10~$\mu$m flux change comes from pulsation,
then we would expect a more than 10\% flux change around 5--7~$\mu$m, which was not observed. 
\label{fig:dTd}}
\end{figure*}

As has been explored by \citet{Reach2009}, there are many different
flavors of disk models that can fit the G29-38's SED including a very
elaborate mineralogical model. Generally, there are two major parts in
the models for white dwarf disks: a component that contributes
most of the featureless mid-infrared flux, and a component that is
responsible for the solid-state feature. We used the geometrically
thin and optically thick disk model (same as for SDSS~J1228+1040 in
Section~\ref{sec:wd1226model}) to represent the featureless component.
The prominent solid-state features in the observed mid-infrared
spectra most likely come from the dust emission in an optically thin
part of the disk, which could be the part of a wrapped disk (e.g., GD
362, \citealt{Jura2007a}) or the outer part of a wedge-like disk
\citep[e.g.][]{Reach2009}. Ideally, one should use a radiative
transfer model to self-consistently compute the disk model for both
parts; however, such a model has many parameters that current,
non-simultaneously obtained multi-wavelength data cannot constrain. A
full exploration of various parameters using a radiative transfer
model will be presented in a future study. Our main goal here is to
understand the temporal change of the 10~$\mu$m feature. Therefore, we
model the G29-38 disk as a two-component disk model. 

We assume that the optically thin part of the disk is the topmost and
outermost layer of the optically thick disk. Both the optically thin
and thick disks share the same inner radius while their outer radii
can be different. A larger outer radius for the optically thin disk (hence cooler material) is
necessary to account for the mid-infrared flux. We assume the dust is sublimated to
gas when its thermal equilibrium temperature reaches the dust
sublimation temperature. In this case, the flux contribution from
these grains is set to zero. The dust in the innermost region of
white dwarf disks can reach 2500--3000~K, which is generally higher
than the same material in a protoplanetary disk due to the presence of
metallic gas \citep{RafikovGarmilla2012}. We set the dust sublimation
temperature to be 2000~K for silicates, and compute thermal equilibrium
dust temperatures for a range of grain sizes and their resultant
emission using the program developed for modeling debris disks around
main sequence stars \citep[e.g.][]{Su2015}. Since the exact mechanism
that creates the dust grains in white dwarf disks is unknown, we adopt
a standard particle size distribution with a power-index of $-$3.5
resulted from collisional cascades \citep{Dohnanyi1969}. Silicate
grains larger than a few $\mu$m in size contribute little to the
strength of the 10~$\mu$m feature, therefore we can only constrain the
largest grain size that is required to fit the shape of the
feature. Larger grains will contribute the underlying continuum, which
are part of the optically thick disk and remain unconstrained under
the assumption of two component model. There are four free parameters
for the optically thin disk: the outer radius of the disk, the minimum
and maximum gain sizes, and the total dust mass (assuming a density of
3.5~g~cm$^{-3}$). 

Given the number of free parameters there is a range of disk
parameters that can fit the G29-38 spectrum satisfactorily. By using
olivine grains (Mg$_{0.4}$Fe$_{0.6}$SiO$_4$,
\citealt{DorschnerHenning1995}) and the low-state of the stellar
temperature (T$_\mathrm{wd,low}$=11015~K), we are able to obtain a very good match in the 10 $\mu$m
feature for the 2004 epoch, and a reasonable fit in the mid-infrared
SED; however, the fit in the 20~$\mu$m region is less good, as shown
in the left panel of Fig.~\ref{fig:dTd}. One possibility for this poor
match is that the 2006 IRS LL spectrum was obtained at a different
epoch from the 2004 SL spectra. A more likely explanation is that the
optical constants for the pure olivine measured in the laboratory are
not exactly matched to the properties of astronomical material. We
will explore different grain composition in Section
\ref{sec:g29-38disk}. It is worth pointing out that in the
two-component model, the optically thin part also contributes to the
continuum flux between 3--8~$\mu$m. For SDSS~J1228+1040, due to the
lack of longer wavelength data, we did not consider this component. The optically thin component is likely to be present due to the high IRAC [8.0] and MIPS fluxes. If such a component is common in dusty white dwarfs, then
it could potentially explain the changes
observed in SDSS~J1228+1040 and SDSS~J0959$-$0200. Future high quality
infrared spectroscopic observations will be crucial to studying the
disk structure around white dwarfs.

We now explore whether the change in the 10~$\mu$m feature could be
caused by the change of the stellar heating. With our base low-state
model parameter determined, we re-compute the dust equilibrium
temperatures using the high stellar temperature (T$_\mathrm{wd,high}$ = 11465 K), and derive the output
SED by using the same disk parameters that fit the 2004
spectrum. Increasing the heating power does increase the overall disk
emission because the grains are at slightly higher temperatures, and
the changes are slightly different between the two parts of the disk. As shown in the right panel of Fig.~\ref{fig:dTd}, the net effect is that the change from stellar pulsation is wavelength dependent. With the fixed disk
parameters (location, grain sizes and mass), we expect the
pulsation-induced change in the disk emission to be higher at 2--7
$\mu$m region than that of the 10 $\mu$m region. The maximum pulsation
induced variability ($\approx$15\%) occurs at 3~$\mu$m, while it is
mostly below 10\% at other wavelengths. We also test a more extreme
case with a large temperature swing of $\pm$1000~K for the star. The
change is more dramatic (a larger flux ratio overall) in this case,
but the relative wavelength-dependent trend remains the same. From
these calculations, we conclude that stellar pulsation is likely to be
responsible for the low level variability observed in the K$_\mathrm{s}$ and IRAC
bands (discussed in Sections~\ref{sec:spitzerphot} and
\ref{sec:jhkphot}). However, the same mechanism is unlikely to explain
the flux change around 10 $\mu$m because we see no flux change in the
in the IRS 5--7~$\mu$m region between all three IRS epochs. In
addition, the integration times of the IRS SL spectra are 366~sec, a
factor of two longer than the pulsation period of 190~sec found in the
infrared light curve. As a result, any pulsation effect on heating the dust is
likely to be averaged out for the IRS observations. Therefore, the
change in the 10 $\mu$m feature most likely comes from the intrinsic change of
the disk parameters in the optically thin part of the disk.

\subsubsection{Variability from an Increase of Dust Mass\label{sec:g29-38disk}}

\begin{figure*}
\gridline{\fig{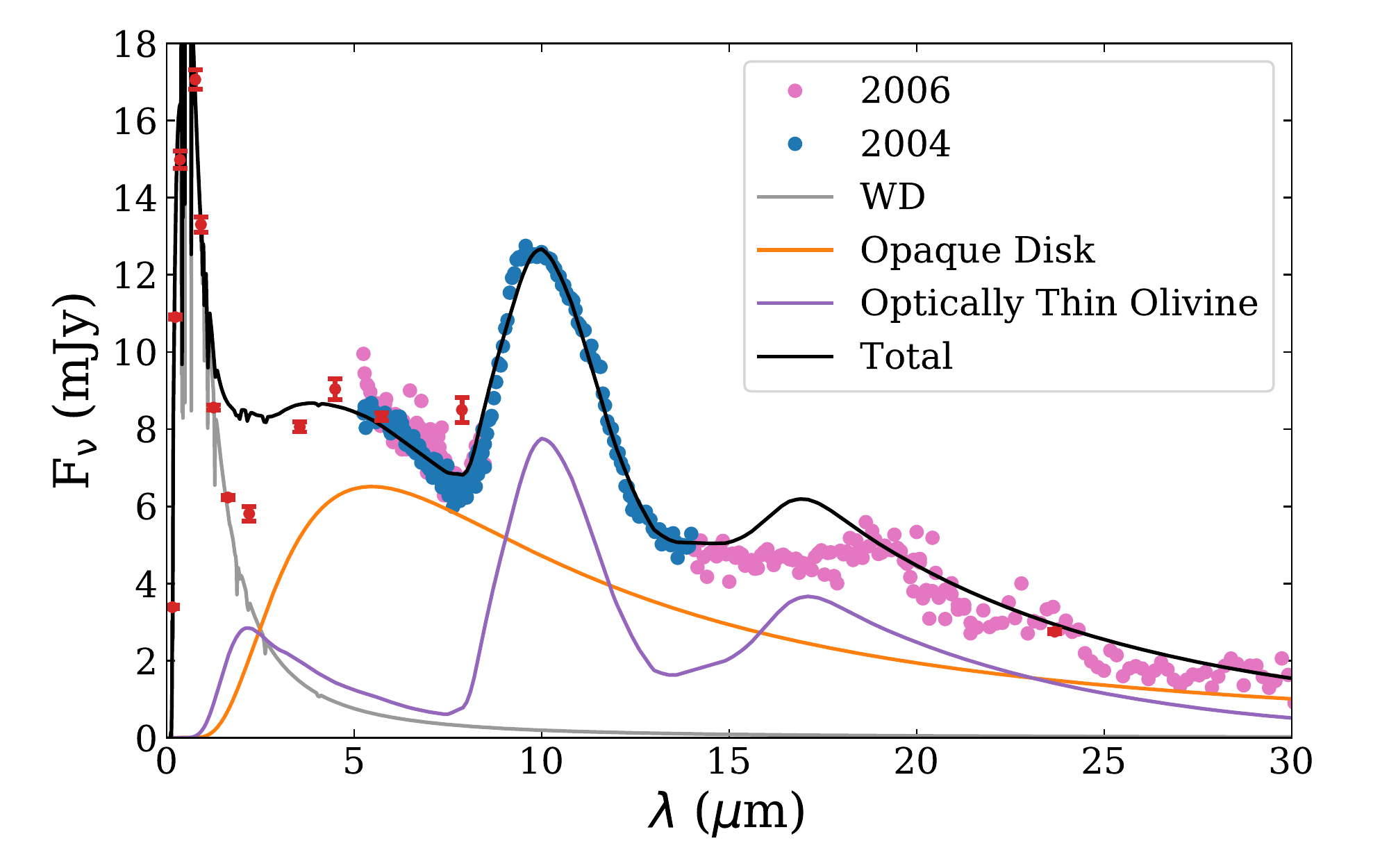}{0.5\textwidth}{}
\fig{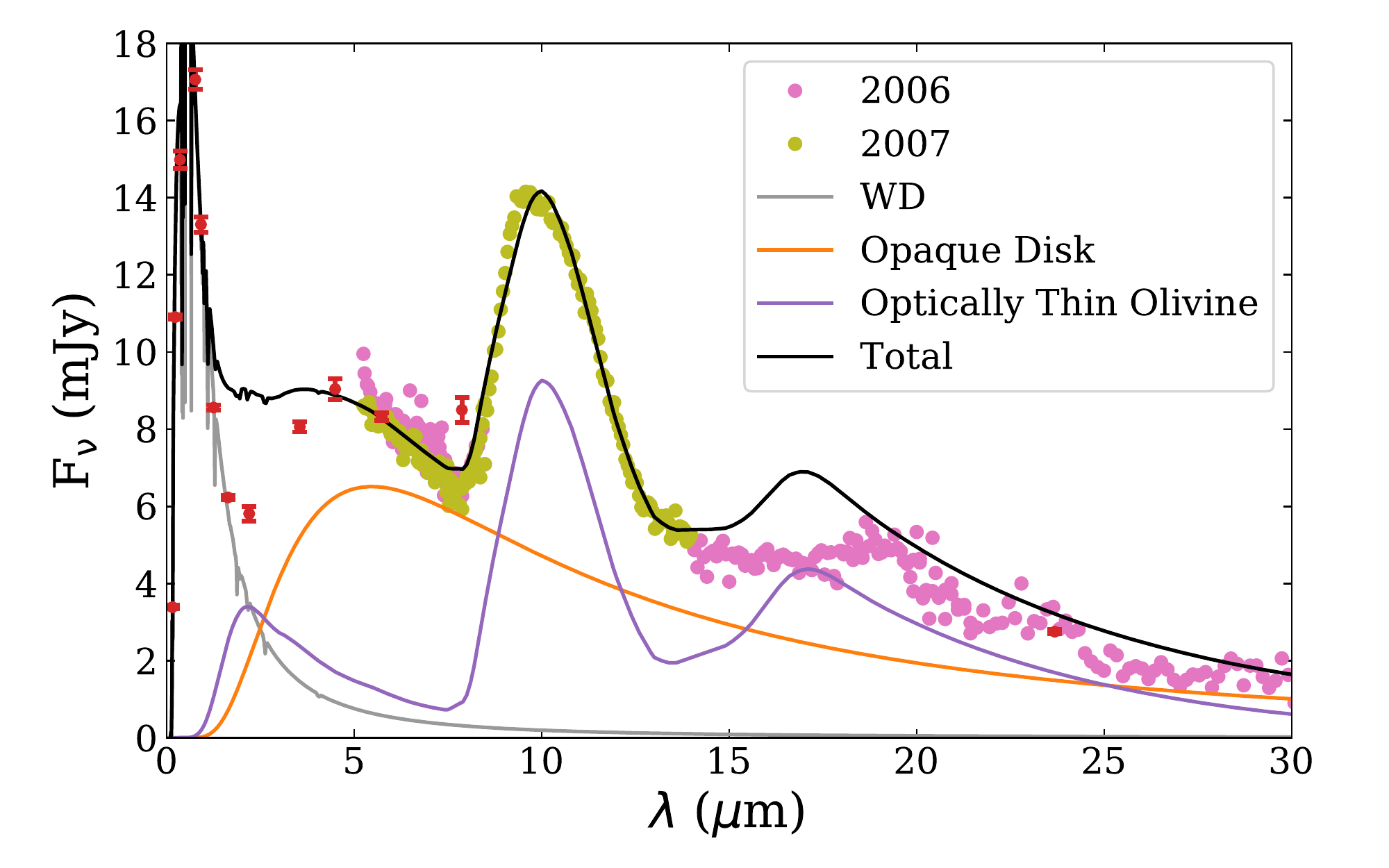}{0.5\textwidth}{} }
\gridline{\fig{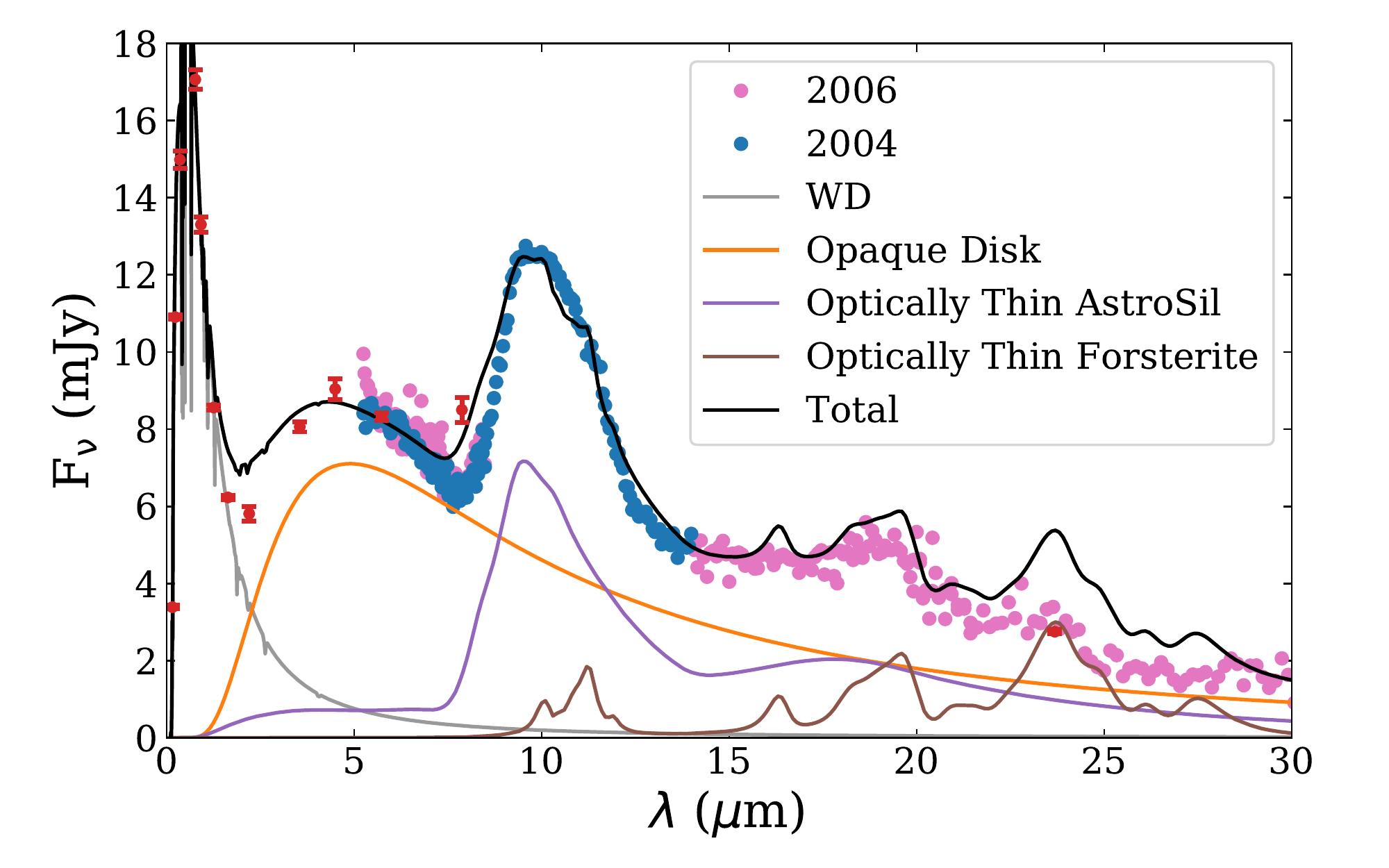}{0.5\textwidth}{}
\fig{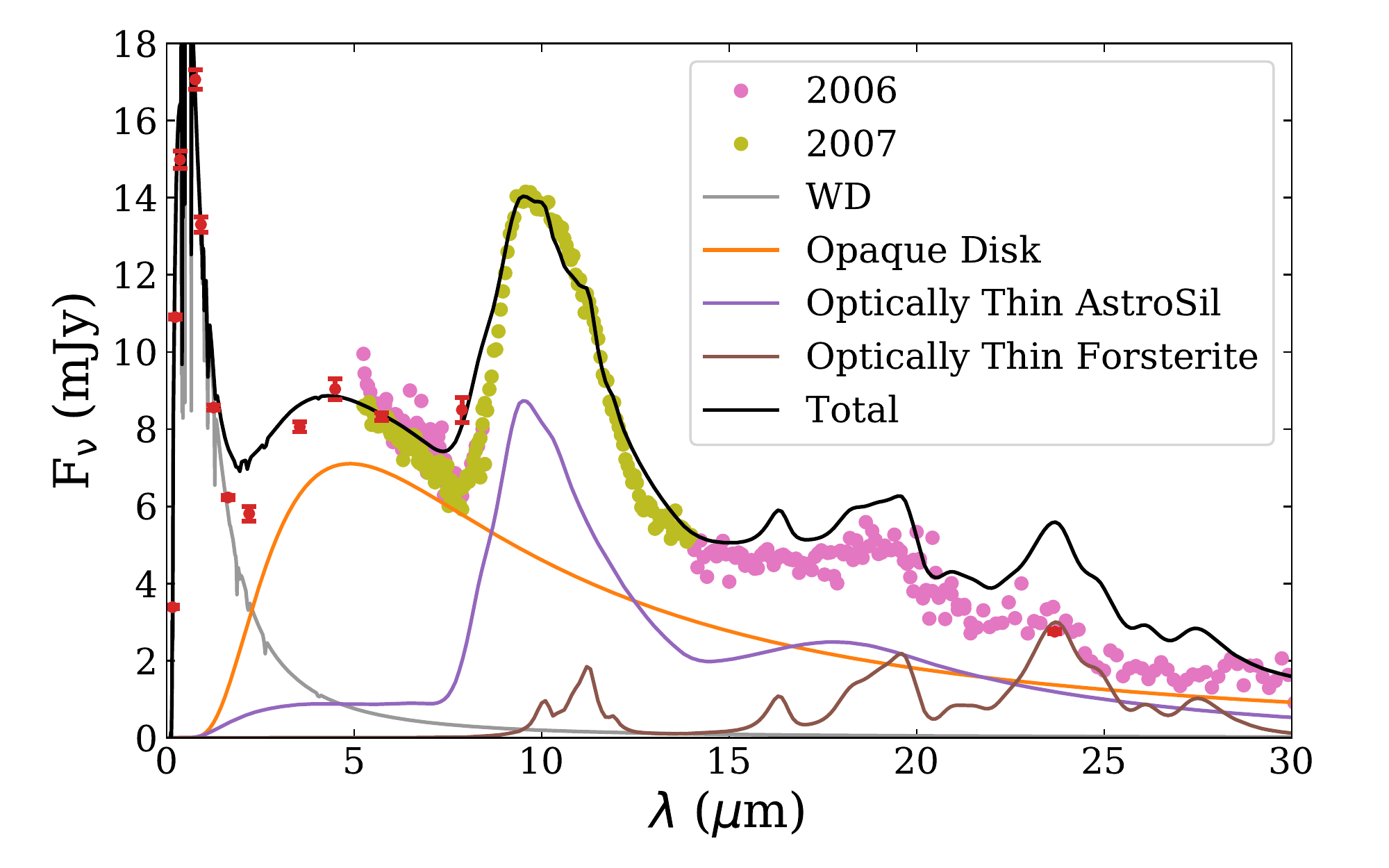}{0.5\textwidth}{} }
\caption{Two-component disk models of G29-38 for fitting the {\it
Spitzer}/IRS data. The top row shows models using olivine composition,
and the bottom shows the ones using a combination of astronomical
amorphous and crystalline silicates. The difference between the 2004
and 2007 IRS data can be explained by an increase of dust mass.
\label{fig:g29-38_sedmodels}}
\end{figure*}

We now proceed to explore the possible changes in the disk parameters
that could be responsible for the 10~$\mu$m feature variation. Since
the change of stellar effective temperature is unlikely to be the main
reason, we fix the stellar effective temperature to be at the
average temperature of 11240 K for the rest of our modeling for
simplicity. In addition to the olivine grains used in the previous
section, we also consider astronomical silicates
\citep{LaorDraine1993} for the grain composition. Similarly, we first
derive the model parameters using the 2004 IRS spectrum for both
compositions. Using the astronomical silicates, as shown in
Fig.~\ref{fig:g29-38_sedmodels}, the 10 $\mu$m feature shape is not a
good match at all (a shift in the peak of the feature), but the fit to
the 17 $\mu$m region is much better compared to the one using olivine
grains. In fact, adding a small amount of crystalline silicates
(forsterite, Mg$_2$SiO$_4$, \citealt{Koike2003}) improves the match to
the red side of the 10 $\mu$m feature and the overall features in the
IRS LL spectrum, consistent with the finding by \citet{Reach2009}. 
Satisfactory fits can be achieved by using either olivine or astronomical silicates with a small amount of forsterite. 

As discussed in Section~\ref{sec:g29-38model}, there is a wide range
of parameters that can give reasonable fits to the G29-38 spectrum. We
are mostly interested in the cause for the change in the 10~$\mu$m
feature, so we deliberately fix most of the parameters in the models,
and only vary the amount of dust in the optically thin disk when
fitting the 2007 epoch of the IRS spectrum. The parameters used for
the fits are given in Table \ref{tab:g29-38fit}.  As shown in
Fig.~\ref{fig:g29-38_sedmodels}, the difference between the 2004 and
2007 spectra can be explained by an increase of $\approx$20\% in the
dust mass for both compositions. We further explore whether the change
in grain sizes in the optically thin disk could explain the difference
seen in the two epochs. Within the uncertainty, the 2007 IRS spectrum
is also consistent with a model using a slightly smaller size range of
olivine grains and an increase of dust mass by 5\%. Under the
two-component disk model, we conclude that the most likely cause of
variability in the IRS spectra is due to an increase (5--20\%) in the
mass of small grains.

Note that the dust mass derived from our optically thin disk model is
lower than the value derived by \citet{Reach2009}, where a total of
2$\times10^{19}$g is needed with a grain size of 0.1--10 $\mu$m. In
addition to the small difference in grain sizes, the majority of the
difference comes from the fact that the mass we estimate does
not include the optically thick part of the disk, where its
contribution was accounted for as the emission from the carbon-like grains
in \citet{Reach2009}.

\begin{deluxetable*}{llccccc}
\tablecaption{ Two-Component Disk Parameters for G29-38 \label{tab:g29-38fit}}
\tablewidth{0pt}
\tablehead{
\colhead{Component} &  \colhead{Description}  & \multicolumn{2}{c}{Olivine composition} & & \multicolumn{2}{c}{Astronomical Silicates$^1$}
}
\startdata
opaque disk   &\ \ \ \ \ \ inner radius R$_\mathrm{in}$      & \multicolumn{2}{c}{10 R$_{wd}$}  & & \multicolumn{2}{c}{8 R$_{wd}$}  \\
	& \ \ \ \ \ \ outer radius R$_\mathrm{out}$ & \multicolumn{2}{c}{28 R$_{wd}$}  & & \multicolumn{2}{c}{25 R$_{wd}$} \\
              &\ \ \ \ \ \ inclination $i$     & \multicolumn{2}{c}{30\arcdeg}       & & \multicolumn{2}{c}{30\arcdeg}   \\
optically thin disk    &\ \ \ \ \ \  grain sizes  & \multicolumn{2}{c}{0.5--5.0 $\mu$m} & & \multicolumn{2}{c}{0.1--3.0 $\mu$m}   \\
              &\ \ \ \ \ \  outer radius    & \multicolumn{2}{c}{167 R$_{wd}$}     & & \multicolumn{2}{c}{834 R$_{wd}$} \\
              &\ \ \ \ \ \  dust mass       & \underline{2004 epoch} & \underline{2007 epoch} & & \underline{2004 epoch} & \underline{2007 epoch} \\
              &                  &   1.1$\times$10$^{18}$ g  &   1.4$\times$10$^{18}$ g          & &  4.0$\times$10$^{18}$ g  &  4.8$\times$10$^{18}$ g    \\
\enddata
\tablenotetext{1}{with a small amount (1.6 $\times$ $10^{18}$ g) of sub $\mu$m forestite grains at a fixed dust temperature of 250 K.}
\end{deluxetable*}

\section{Implications \label{sec:interp}}

It is exciting that we are detecting temporal variation of dust disks around white dwarfs. SDSS~J1228+1040, together with SDSS~J0959$-$0200, seem to belong to the same category. Their [3.6] and [4.5] fluxes dropped by 20-30\% within a few years and then mostly remain the same afterwards. For G29-38, the 10~$\mu$m feature has increased by 10\% within three years while the 5--7~$\mu$m fluxes remained the same. Because of the limited coverage, we do not know whether it is a sudden or gradual drop in the infrared flux or a temporal increase followed by a drop. We also do not know yet the long term trend of these variations. In addition, SDSS~J1228+1040 and SDSS~J0959$-$0200 both display significant amounts of circumstellar gas. 

For G29-38, the increase of 10~$\mu$m flux is most likely caused by an increased dust mass of $\sim$ 10$^{17}$g in the optically thin component. A continuous change between the two IRS observations (2004 Dec 8 and 2007 Aug 4) would correspond to a dust production rate of 10$^9$ g s$^{-1}$, which is comparable to the mass accretion rate of 7 $\times$ 10$^8$ g s$^{-1}$ observed in the atmosphere of G29-38 \citep{Xu2014}.

For SDSS~J1228+1040, under the opaque disk assumption, the change in the 3--8~$\mu$m flux can be explained by a change in the disk inclination or a decrease of disk surface area by 8--48\%. We can estimate the lower limit of the mass of the opaque disk to be $\pi$ (R$_\mathrm{out}^2$ - R$_\mathrm{in}^2$) $\times$ 2$h$ $\times$ $\rho$ $\sim$10$^{23}$~g assuming a scale height $h$ of 1~cm and a density $\rho$ of 3~g~cm$^{-3}$. Such a decrease in mass over the two IRAC observations separated by 7 years would require a mass loss rate of 4$\times$10$^{13}$ g s$^{-1}$ -- 2$\times$10$^{14}$ g s$^{-1}$. This rate is significantly higher than the mass accretion rate of 6~$\times$~10$^8$ g s$^{-1}$ derived from its atmospheric pollution \citep{Gaensicke2012}.

Here, we explore some mechanisms that could lead to the observed infrared variability.

\subsection{Tidal Disruption Event}

As discussed in \citet{Jura2008}, tidal disruption of a massive minor planet could disrupt the pre-existing disk. Because the incoming body is likely to have a different orbital inclination, a new dust disk will eventually be formed at a different orbital inclination. The final infrared flux could be either higher or lower than the previous one. In this scenario, a significant amount of dust will be evaporated to produce circumstellar gas, which could explain the presence of the calcium infrared triplet observed around SDSS~J1228+1040 and SDSS~J0959$-$0200. Because the emission lines are optically thick, it is possible for the total amount of gas to change without the strength of the emission line changing. However, from the smooth change of the calcium infrared triplet lines of SDSS~J1228+1040 from 2003 to 2015 \citep{Manser2016a}, there is little evidence for a tidal disruption event.

In order for an incoming minor planet to disrupt the pre-existing dust disk, it needs to be significantly more massive than the dust disk, which is estimated to be 10$^{23}$~g. Given that current infrared monitoring of dusty white dwarfs is sparse, the chance of finding variable systems is slim. However, finding at least two infrared variable dusty white dwarfs (SDSS~J1228+1040 and SDSS~J0959$-$0200) out of a total of 43 systems over 7 years gives an upper limit of one major tidal disruption event (mass $>$~10$^{23}$~g, diameter $>$~200~km) every 140 years, which is significantly shorter than the frequency of tidal disruption events derived from dynamical simulations \citep{Veras2016, Mustill2018}. In addition, with such an energetic tidal disruption event, the infrared excess is likely to display a temporary increase, which needs to fit in this already extremely short timescale. 

Apart from these major tidal disruption events, it has been suggested that accretion of small planetesimals ($<$~35~km, $\sim$~10$^{19}$~g) are nearly continuous \citep{Wyatt2014}. The increase of the 10~$\mu$m feature in G29-38 could come from continuous accretion of small planetesimals, if these all became small dust. An increase of 10$^{17}$~g could come from an object of 4~km in diameter.

\subsection{External Perturbation}

\begin{figure}
\plotone{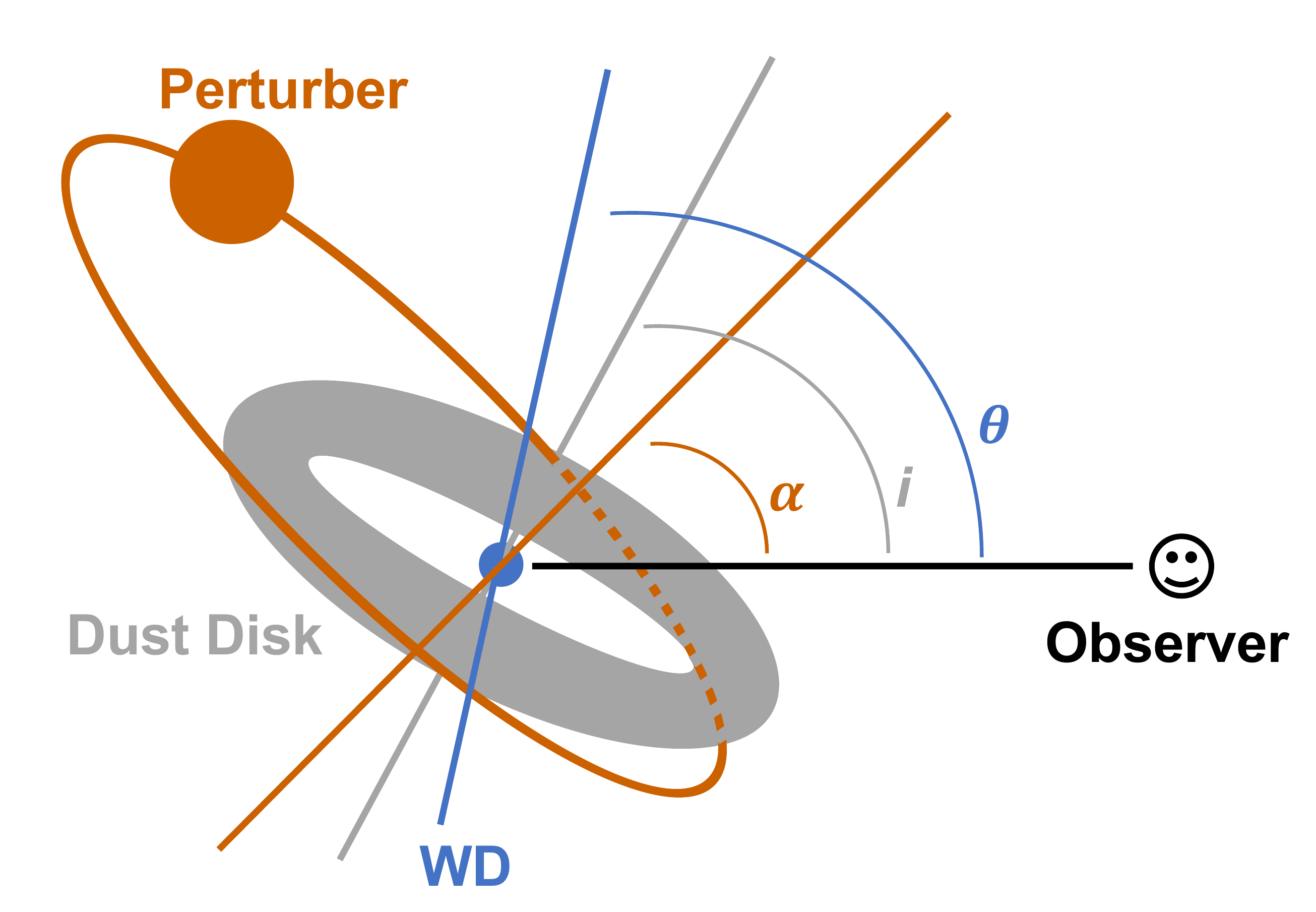}
\caption{A cartoon illustration of a white dwarf, a dust disk, and an unseen perturber. Their orbital planes are not aligned and $\alpha$, $i$, $\theta$ are defined as the angle between our line of sight and the orbital plane of the perturber, the orbital plane of the dust disk, and the white dwarf rotation axis (pulsation axis). Note that the figure is not to scale.
}
\label{fig:illus}
\end{figure}

Here, we explore the response of the dust disk in the presence of an external perturber on an inclined orbit. By studying the pulsation light curve of G29-38, \citet{Montgometry2005} derived the rotation axis (pulsation axis) $\theta$ to be 65$^\circ$ (see Fig.~\ref{fig:illus} for a cartoon illustration), which is comparable to 55$^\circ$ derived from comparing the amplitudes of the harmonics \citep{Thompson2010}. The white dwarf rotation axis might not be aligned with the axis $i$ of the opaque disk, which is derived to be 30$^\circ$\footnote{The inclination of the opaque disk also depends on the contribution of the optically thin component in 3-8 $\mu$m (see Section~\ref{sec:g29-38model}).}. This is not surprising given tidal disruption is a dynamical process and the inclination of the orbits of the minor planets can vary significantly. An external perturber that recently arrived on an orbit which is inclined with respect to the dust disk can perturb the latter and possibly cause the infrared variability. These perturbers are expected to be the catalysts for the pollution of white dwarf atmospheres \citep{DebesSigurdsson2002, VerasGaensicke2015}.

To quantitatively estimate the effect of an external perturber on the dust disk, we ran a basic suite of $N$-body simulations for the SDSS~J1228+1040 system. We modeled the disk as a series of 5 coplanar circular rings each containing 30 massless particles uniformly distributed in azimuth. The location of the rings were at $20, 40, 60, 80$ and $100 R_\mathrm{wd}$ away from the center of the white dwarf. We carried out the simulations with a modified version of the {\it Mercury} integration package \citep{Chambers1999}, which allowed us to incorporate the effects of general relativity.  This exercise is in the same spirit as that in \citet{Manser2016a} except here the perturber's orbital plane is not aligned with the dust disk ($\alpha$ $\neq$ $i$ in Fig.~\ref{fig:illus}). 

\begin{figure}
\epsscale{1.15}
\plotone{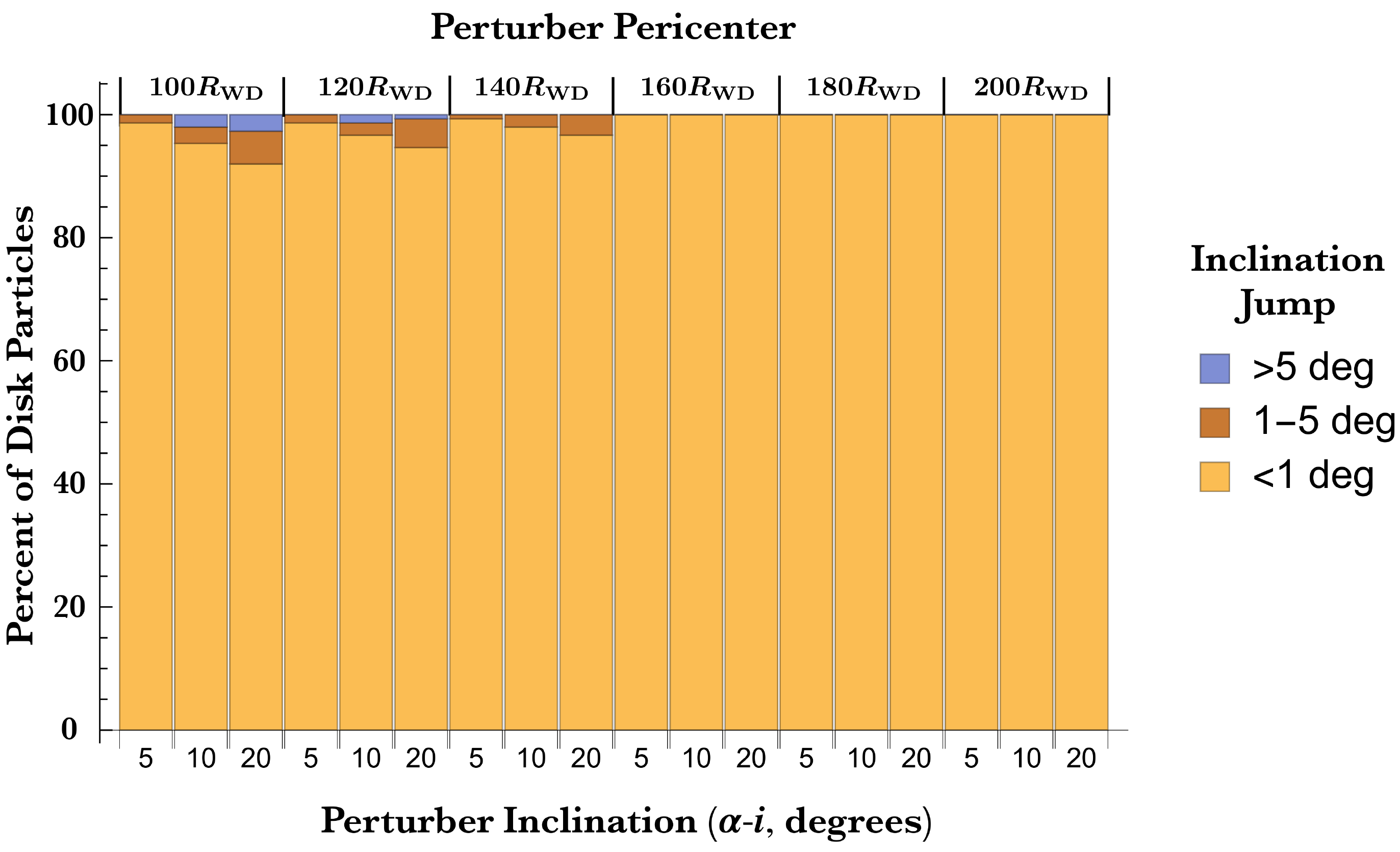}
\caption{Excitations of disk particles at different orbital radii as a function of perturber orbital pericenter and perturber inclination $| \alpha - i |$. This simulation is for one pericenter passage of a $10M_{\rm J}$ planet with a semimajor axis of 5 AU. The plot suggests that even in the most optimistic case, only a small part of the disk can reach an inclination change larger than 1 degree.
\label{fig:nbodysim}}
\end{figure}

We found that the degree of perturbation is strongly dependent on the perturber's mass and orbital properties (semimajor axis, eccentricity, inclination). We focus on an extreme case with a 10~$M_\mathrm{J}$ perturber at a semi-major axis of 5~au while the pericenter of the perturber's orbit (ranging from 100~R$_\mathrm{wd}$ to 200~R$_\mathrm{wd}$) and orbital inclination ($| \alpha - i |$ ranging from 5$^\circ$ to 20$^\circ$) are taken at different values. The result from one pericenter passage is shown in Fig.~\ref{fig:nbodysim}.  The disk is perturbed the most when the perturber has a pericenter closest to the disk edge with a large inclination angle. However, even in the most optimistic case, only a small fraction of the disk can be perturbed by more than 1$^\circ$ after one pericenter passage, which is not sufficient for a 4$^\circ$ inclination change required to reproduce the change of the infrared excess. 

We consider in our simulation a typical pericenter passage; in reality, the disk would be perturbed every time the perturber passes the pericenter. In addition, our simulations show that with a perturbing object, the disk will no longer remain flat, which is also suggested by numerical studies of dust disks \citep{KenyonBromley2017a}. This change could reduce the optical depth of the dust disk and correspondingly increase the dust temperature, leading to grain sublimation and a drop in the infrared luminosity. Exploring this scenario would require $N$-body simulations over multiple pericenter passages and is beyond the scope of this work.

\subsection{Runaway Accretion}

The infrared variability of SDSS~J1228+1040 and SDSS~J0959$-$0200 is very similar and they both display gas emission lines from the calcium infrared triplet. It has been suggested that when there is a strong coupling between the dust and gas, runaway accretion could occur and it can lead to a higher accretion rate than Poynting-Robertson drag can support \citep{Rafikov2011b,Metzger2012}. The evolution of the disk is significantly different from those without gas drag.

 For SDSS~J1228+1040, the mass accretion rate supported by Poynting-Robertson drag is $\sim$~10$^9$ g s$^{-1}$. During runaway accretion, the peak value can reach $\sim$~10$^3$ higher than the rate from Poynting-Robertson drag \citep{Metzger2012}. This higher value is still a bit lower than the derived mass loss rate of the dust disk (10$^{13}$ - 10$^{14}$~g~s$^{-1}$). For runaway accretion to occur, there are two main criteria: I. the dust disk is massive enough ($\gtrsim$ 10$^{22}$ g); II. there is a build-up of gas due to sublimation occurring faster than the rate that gas is removed by viscous diffusion \citep{Rafikov2011b}. Both SDSS~J1228+1040 and SDSS~J0959$-$0200 have strong infrared excesses, which likely originate from a massive disk. Criterion II requires a strong gas-solid coupling factor and/or a weak gas viscosity. Although there are a lot of uncertainties in these parameters, it could be satisfied for white dwarf disks \citep{Metzger2012}.

Runaway accretion has not been directly observed in white dwarf disks. An indirect line of evidence is that some of the helium-dominated white dwarfs with circumstellar gas (e.g. WD~J0738+1835, Ton~345) have the highest mass accretion rates of all polluted white dwarfs, which is expected from runaway accretion. Once runaway accretion starts, the solid disk will be dissipated in a very short amount of time. Correspondingly, the infrared flux will continue to drop and the mass accretion rate would increase substantially \citep{Metzger2012}. So far, there is no evidence of continuing drop in the infrared flux nor an increase of absorption line strength in the atmosphere \citep{Manser2016a}. Future monitoring would be crucial in assessing the runaway accretion scenario.

\section{Conclusion \label{sec:conclusion}}

In this paper, we presented infrared variabilities of two white dwarfs with dust disks, SDSS~J1228+1040 and G29-38. For SDSS~J1228+1040, the IRAC [3.6] and [4.5] fluxes dropped by 20\% within 7 years and remained the same afterwards. The general behavior is very similar to the flux drop around another dusty white dwarf WD~J0959$-$0200 \citep{XuJura2014} with a further similarity being the appreciable amounts of circumstellar gas around both objects. The flux drop can be explained by either an increase of an inner disk radius, a decrease in the outer disk radius, or a change in the disk inclination, assuming the excess comes from an opaque dust disk.

G29-38 appears to represent a different kind of infrared variability and the flux of the 10~$\mu$m feature has increased by 10~\% within 3 years while the 5--7~$\mu$m flux remained the same. We presented a two-component disk model to fit the infrared spectra, and concluded that the change is unlikely to be related to the photospheric pulsation of G29-38 with a static disk. We propose that the most likely cause is an increase in the mass of small grains in the optically thin component.

To explain the observed infrared variability, we explored several scenarios, including tidal disruption events, external perturbation, and runaway accretion. Although continuous tidal disruptions of small planetesimals could explain the increased dust mass in G29-38, no satisfactory scenarios can explain the sudden drop of infrared flux for SDSS~J1228+1040 and SDSS~J0959$-$0200. 

Looking forward, a self-consistent radiative transfer disk model would be valuable in constraining white dwarf disk parameters. In the future, continued photometric monitoring in the infrared and high quality infrared spectroscopy from the {\it James Webb Space Telescope} will be crucial in constructing a complete picture of dust disks around white dwarfs.

{\it Acknowledgements} The authors would like to thank the late UCLA Professor Michael Jura for his inspirations and support to study white dwarf disks. We thank S. Kleinman and A. Nitta for helpful discussions about white dwarf pulsation. A.\,B. acknowledges a Royal Society Dorothy Hodgkin Fellowship. J.\,O. acknowledges financial support from the ICM (Iniciativa Cient\'ifica Milenio) via the N\'ucleo Milenio de Formaci\'on Planetaria grant, from the Universidad de Valpara\'iso, and from Fondecyt (grant 1180395). D.\,V. gratefully acknowledges the support of the Science and Technology Facilities Council via an Ernest Rutherford Fellowship (grant ST/P003850/1). T. G. W. wishes to acknowledge funding from a STFC studentship. The research leading to these results has also received funding from the European Research Council under the European Union's Seventh Framework Programme (FP/2007-2013) / ERC Grant Agreement n. 320964 (WDTracer).

This work is based in part on observations made with the {\it Spitzer Space Telescope}, which is operated by the Jet Propulsion Laboratory, California Institute of Technology under a contract with NASA. This publication makes use of data products from the Wide-field Infrared Survey Explorer (WISE), which is a joint project of the University of California, Los Angeles, and the Jet Propulsion Laboratory/California Institute of Technology, funded by the National Aeronautics and Space Administration. This publication also makes use of data products from the Near-Earth Object Wide-field Infrared Survey Explorer (NEOWISE), which is a project of the Jet Propulsion Laboratory/California Institute of Technology. NEOWISE is funded by the National Aeronautics and Space Administration.

\end{CJK}

\software{IRAF \citep{IRAF1,IRAF2}, Mercury \citep{Chambers1999}, Matplotlib \citep{Matplotlib}}

\end{document}